\documentclass[aps,prb,twocolumn,groupedaddress,amsmath,amssymb,showpacs]{revtex4-1}

\usepackage{graphicx,graphics,color,epsfig}
\usepackage{amsmath}
\usepackage{amssymb}
\usepackage{amsfonts}
\usepackage{amsfonts}
\usepackage{epstopdf}
\usepackage{bm}
\usepackage{times,xspace}
\usepackage{color}
\usepackage[utf8]{inputenc}
\usepackage{float}
\usepackage{subfigure}

\def\beq{\begin{eqnarray}}\def\eeq{\end{eqnarray}}
\def\be{\begin{equation}}\def\ee{\end{equation}}

\begin{document}

\title{Accessing different topological classes and types of Majorana edge states in coupled superconducting platforms using perturbations}

\author{Sayonee Ray}
\email{srayatunm@unm.edu}
\affiliation{Center for Quantum Information and Control (CQuIC),
Department of Physics and Astronomy, University of New Mexico}

\author{Subroto Mukerjee}
\affiliation{Department of Physics,
Indian Institute of Science, Bangalore, India}

\author{Nayana Shah}
\affiliation{Department of Physics, Washington University, St. Louis} 
\date{\today}


\begin{abstract}

The study of topological classes and their associated edge states has been of ongoing interest. In one dimension, the standard platform of these studies has been the conventional Kitaev wire and its realizations. 
In this work, we study the edge states in coupled p-wave platforms in 1D, in the presence of experimentally relevant perturbations, like a Zeeman field and s-wave SC. 
Firstly, we show that the unperturbed coupled p-wave setup by itself can have two types of Majorana edge states, depending on the value of the effective onsite potential. 
We show that additional components like Zeeman field and s-wave term can cause transitions to different symmetry classes, both topologically trivial or non-trivial, and change the nature of these edge states. In the presence of the perturbations, we show that there are 3 symmetry classes when the effective p-wave pairing is equal between the spin species, and 6 for the second kind, when the pairing differs by a phase $\pi$ between the two. Some of these classes are topologically non-trivial. Further, we explore the nature of subgap states when we have a junction between two such topological setups and their corresponding behaviour with the phase of the p-wave order parameter. Our work provides a theoretical framework of the different ways to get non-trivial topological classes in coupled p-wave nanowire setup, using experimentally feasible perturbations, and the nature of subgap states across junctions of these platforms.
\end{abstract}
\maketitle

Majorana fermions (MF) and topological superconductors have gathered massive interest due to their non-Abelian exchange statistics and have lead to the idea of low decoherence topological quantum computation~\cite{kitaev,stern,fu-kane,sau-sarma,oreg,nayak}.  One of the first theoretical models of a topological SC was the 1D Kitaev model, a one-dimensional spinless p-wave superconductor (SC), which supports isolated Majoranas at the edges~\cite{kitaev}. Majoranas also appear at the points where transition occurs between a topological and a non-topological phase. For example, in the 1D Kitaev model, if the onsite potential $\mu$ has a spatial dependence, then between the regions $|\mu|>2t$, (where, $t$ is the hopping amplitude) and $-t<\mu<t$, two MFs appear at the transition point where the gap closes. 

There is still considerable interest in realizing Majorana fermions experimentally in different geometries and hybrid structures~\cite{fu, qi, shabani1, shabani2}, especially from the topological quantum computation community.  Most widely studied platforms have been systems with strong Rashba spin-orbit coupling (SOC)
in the presence of an external tunable magnetic field~\cite{fu-kane, sau-sarma, oreg, grosfeld, pientka, cook, mohanta, brouwer}. However, recently, there has been significant efforts in exploring more unconventional yet viable platforms which can host Majorana bound states (MBS), for example, systems which do not require Rashba SOC nor external magnetic field~\cite{kim, yazdani, neupert, klinovaja, kjaergaard, vazifeh}. In a recent experimental work~\cite{shabani2}, it was shown that a hybrid semiconductor-superconductor nanowire on the top of a magnetic film in the stripe phase can support the formation of MBS localized at the ends of the nanowire.

The detection and verification of MBS itself is another challenging aspect and has been debated for long. The initial proposals were based on the detection of the zero-bias conductance peaks in nanowires which had the possibility of containing the  trivial zero-energy Andreev bound states~\cite{mourik, das, rokhinson, molenkamp, deng}. In a couple of recent works, observation of $4\pi$ periodic Josephson current was observed across different geometric junctions of topological superconductors, which is another compelling signature of current being carried by single electron processes, contrary to that in conventional superconductors~\cite{yu, calvez, smitha, laroche, wiedenmann}. Even this procedure is often hindered by quasi-particle poisoning, especially in experiments under stationary equilibrium conditions, and one needs to be sure that this $4\pi$-periodicity is not caused by Landau-Zener transitions between topologically trivial subgap states~\cite{schulenborg}. Similar unexplored questions still exist in the physics of topologically trivial and non-trivial subgap states which are relevant to the detection of MF~\cite{schulenborg, annica}.

The existing fault tolerant topological quantum computation scheme use braiding of anyons, satisfying non-abelian exchange statistics, to realize unitary quantum gates~\cite{kitaev, sarma}. In Ref.~\cite{ma}, Ma et al. have shown theoretically that by vortex manipulation in a topological superconductor one can realize quantum gates by using the exchange and braiding operations of MFs. However, in order to implement braiding operations one needs to obtain a stable pair of Majorana bound states (MBS). One way to realize such a system is to have two (or multiple) copies of the Kitaev wire~\cite{flensberg}, each having  p-wave pairing $\triangle_{\uparrow\uparrow}=-\triangle_{\downarrow\downarrow}$ in each spin sector. Based on the discrete symmetries present in the system, anti-unitary  time reversal (TR) $\mathcal{T}$ and particle-hole (p-h) $\mathcal{C}$, and unitary chiral $\mathcal{S}$, this model falls in the topologically non-trivial BDI symmetry class, hosting an integer number of edge states~\cite{schnyder}. In Ref.~\cite{flensberg}, Gaidamauskas {\it et al.} have shown that a pair of TRS nanowires proximity coupled to a superconductor can be driven into a nontrivial topological phase supporting a Kramers pair of MBS at the edge. The exchange of two Kramers pairs of MBS can also constitute a non-Abelian operation in the absence of chiral symmetry~\cite{law}. Additional perturbations on these systems can cause transition to a different symmetry class, which can be topologically trivial.
%

In this work, we focus on the low energy sector of a time reversal (TR) symmetric two-channel quantum wire, proximity coupled to a conventional s-wave SC. In the work by Gaidamauskas et al.~\cite{flensberg}, they derive the low energy effective Hamiltonian and discuss the topological properties of the coupled wire platform, which resembles a spinless Kitaev wire with effective parameters. 
The effective onsite potential $\mu_{eff}$ is a function of the chemical potential of each of the coupled wires, inter-wire and intra-wire superconducting pairing, hopping between the wires, and the external voltage. The effective p-wave parameter is also a function of the spin orbit coupling, due to the proximity coupling to a s-wave superconductor, in addition to the above quantitites. Similarly, an additional effective s-wave term can also be turned on, which is a function of the above mentioned quantitites in the coupled platform. Further details of the derivation is given in the supplemental information of the Ref.~\cite{flensberg}.  In this work, we show that one can access different topological phases hosting edge MBS pairs, using multiple combinations of Zeeman fields and additional s-wave term in this coupled p-wave platform. We classify all such possible combinations into different symmetry classes according to the tenfold scheme~\cite{schnyder}. 

Further, we study the type of edge states in this platform for different values of the onsite potential $\mu_{eff}$. Since $\mu_{eff}$ in this model is not the chemical potential of a single p-wave superconductor, we cannot make the usual assumption that it is the largest energy scale. However, we do restrict ourselves in the topological phase, which for this case is for all $\mu_{eff}>0$. In this paper, from here onwards we will refer to the effective parameters in the model, $\mu_{eff}$ as $\mu$, $\triangle_{eff}$ as $\triangle$ to avoid confusion. The edge states in this effective model are MBS, and have an oscillating part along with a decaying part for $\mu>1/2$  (similar to the form $\sim e^{-x} \sin(x)$). For $0<\mu<1/2$, the MBS are purely decaying. A similar study was reported by Klinovaja et al.~\cite{jelena2012} in a quasi-one-dimensional nanowire system containing SC and normal sections in weak and strong spin-orbit interaction regimes, where they showed different spatial dependence of MBS depending on whether they are deep in the topological phase or not. 

We have considered two kinds of spinless p-wave pairing with 
$\triangle_{\uparrow\uparrow}=\triangle_{\downarrow\downarrow}$ and $\triangle_{\uparrow\uparrow}=-\triangle_{\downarrow\downarrow}$, which can arise in this platform depending on the intra-wire and inter-wire paring between the coupled wires. The first kind of Hamiltonian is realizable in the low energy sector of a time reversal (TR) symmetric two-channel quantum wire, proximity coupled to a conventional s-wave SC, and is found to support a Kramer's pair of Majorana bound states (MBS) in the topological phase~\cite{flensberg}. They showed that transition between different topological phases, hosting two, one or zero number of MBS, can be effected by applying a magnetic field perpendicular to the spin-orbit direction. It was shown by Tewari {\it et.al}~\cite{tewari1} that MBS in a TR symmetric 1D p-wave chain are topologically robust to perturbations which are TR symmetry breaking, like the magnetic field. It was identified that with perturbations such systems belong to the BDI symmetry class, whereas the TR symmetric Hamiltonian can be characterised as both BDI or DIII. 
We show that 1D spinless p-wave SC lies in the BDI class, and not in the DIII class, which can be seen by redefining the TR and p-h operators to a more general form as we discuss below in Appendix~\cite{schnyder}(for standard spinful systems, the TR operator is $i \sigma_y\mathcal{K}$).  Such systems have $Z$ topological invariant and an integer number of edge modes, which is reflected by the MBS doublets at the edges~\cite{flensberg,tewari1,tewari2}. 
We show that adding perturbations like the s-wave and Zeeman fields will induce transitions from one topological phase to another. There are 3 symmetry classes for the first kind and 6 for the second. Additionally, we analyze the subgap states across junctions of such hamiltonians belonging to different topological classes and their behaviour with the phase of the p-wave order parameter. This is relevant for studying transport properties across junctions of topological phases in coupled nanowire platforms. Finally, we use a low energy effective Hamiltonian (for small k we can assume, $\cos{k}\sim (1-k^2/2)$ and $\sin{k} \sim k $)  to study the behaviour of the subgap states and differentiate between the two cases, with $\mu>1/2$ (deep in the topological phase) and $\mu<1/2$, in the BDI class.
%

The paper is arranged as follows. In Sec.~\ref{spinfullbdg}, we discuss the spinful BdG Hamiltonian and the kinds of p-wave pairing that can arise in these systems. In Sec.~\ref{symmclass}, we study the symmetry classes and the MBS solutions that can arise in the coupled p-wave platform.
Using the tenfold scheme, we have classified all possible combinations of Zeeman field and s-wave order parameter (based on the TR, p-h and chiral symmetries) that can keep the system topologically non-trivial. This section is divided into two. In Sec.~\ref{Sec1}, we focus on the case for which $\triangle_{\uparrow\uparrow}=\triangle_{\downarrow\downarrow}$.
In Sec.~\ref{Sec2}, the analysis is for $\triangle_{\uparrow\uparrow}=-\triangle_{\downarrow\downarrow}$. Combinations of perturbations that give rise to a topologically trivial class are enlisted and discussed in Appendix, in Sec.~\ref{appendix}.
We have explored the possibility when both the Zeeman fields ($B_1\hat{\sigma}\tau_z$ and $B_2\hat{\sigma}\tau_0$) are simultaneously present as perturbations~\cite{tewari1,tewari2,flensberg}. Without loss of generality, we have fixed $B_1$ in the $x$-$z$ plane and have considered $B_2$ to be general. In Sec.~\ref{JJ}, we analyze the subgap states across junctions between BDI classes. This is relevant for studying transport properties across junctions of topological superconductors~\cite{nayana1}. The motivation of this section is to study and differentiate the nature of the subgap states between the cases when $\mu>1/2$ and $\mu<1/2$ . Point to note, in our analysis, the onsite potential is not assumed to be the largest energy scale, contrary to the general approach of calculating the Josephson current or subgap states in conventional SC. The onsite potential $\mu$ is an effective parameter of the inter-wire and intra-wire pairing, inter-wire hopping and the chemical potential of the individual wires, and is not the bare chemical potential (or Fermi energy) of the superconducting wire. Relaxing this assumption helps in understanding the behaviour of the subgap states when it is deep in the topological phase, especially as they approach the bulk bands. 

\section{p-wave superconductivity with spin} \label{spinfullbdg}
The general form of a spin-full BdG Hamiltonian with p-wave pairing, in the absence of a magnetic field, is given by:
\begin{eqnarray} \label{pwave_BdG}
H_{BdG} &=& 
\begin{pmatrix}
\frac{\epsilon_{p,\uparrow}}{2} & \triangle_{\uparrow \downarrow}(p) & 0 & \triangle_{\uparrow \uparrow}(p)\\
\triangle^*_{\uparrow \downarrow}(p) & -\frac{\epsilon_{-p,\downarrow}}{2} & \triangle^*_{\downarrow \downarrow}(p) & 0\\
0 & \triangle_{\downarrow \downarrow}(p) & \frac{\epsilon_{p,\downarrow}}{2} & \triangle_{\downarrow \uparrow}(p)\\
\triangle^*_{\uparrow \uparrow}(p) & 0 & \triangle_{\downarrow \uparrow}(k) & -\frac{\epsilon_{-p,\uparrow}}{2}\\
\end{pmatrix}
\end{eqnarray}
in the basis, $\begin{pmatrix}
c^{\dagger}_{p,\uparrow} & c_{-p,\downarrow} & c^{\dagger}_{p,\downarrow} & c_{-p,\uparrow}
\end{pmatrix}$ . 

In the presence of two different species of electrons (say, spin) labelled by $\alpha$ and $\beta$, triplet pairing between them should follow:
$
\triangle_{\alpha \beta}({\bf p}) = -\triangle_{\alpha \beta}(-{\bf p})$, and,
$\triangle_{\alpha \beta}({\bf p}) = \triangle_{\beta \alpha}({\bf p}).
$

With the above properties, some of the possible pairings are: 
\begin{eqnarray} \label{3_pairing} \nonumber
\triangle_{\uparrow \uparrow}({\bf p})&=&\triangle_{\downarrow \downarrow}({\bf p}) \nonumber\\
\triangle_{\uparrow \uparrow}({\bf p})&=&-\triangle_{\downarrow \downarrow}({\bf p})\nonumber\\
\triangle_{\uparrow \downarrow}({\bf p})&=&\triangle_{\downarrow \uparrow}({\bf p})
\end{eqnarray}

The special cases of Hamiltonians with these possible pairings are:
\begin{eqnarray} \label{3_ham} \nonumber
\mathcal{H}_1 &=& \left(\frac{p^2}{2m}-\mu\right)\tau_z + \triangle_0 p\sigma_0 \tau_x \nonumber\\
\mathcal{H}_2 &=& \left(\frac{p^2}{2m}-\mu\right)\tau_z + \triangle_0 p\sigma_z \tau_x \nonumber\\
\mathcal{H}_3 &=& \left(\frac{p^2}{2m}-\mu\right)\tau_z + \triangle_0 p\sigma_x\tau_x
\end{eqnarray}
where, $\triangle_0$ gives the magnitude of the p-wave pairing, $\sigma$ is the Pauli matrix for spin and $\tau$ for particle-hole sector. In the first case, the pairing is between same species with same sign for $\uparrow$ spin and $\downarrow$ spin~\cite{flensberg}, whereas for the second case, they have opposite signs. In the third case, the pairing is of the form $\langle \hat{c}_\sigma({\bf p}) \hat{c}_{-\sigma}(-{\bf p})\rangle$~\cite{KSG}. In this work, we focus on the first and second kinds of p-wave pairing, which can arise in coupled superconducting platforms, as discussed by Gaidamauskas et al. in ~\cite{flensberg}.

\begin{figure}
\centering
\includegraphics[width=8cm,height=4cm]{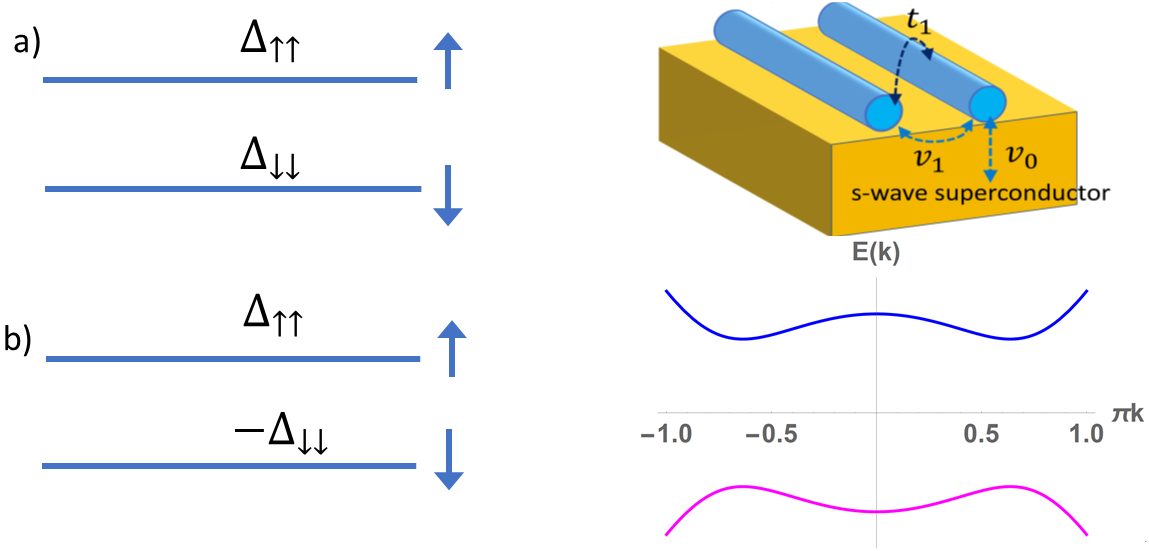}
\caption{A platform with two superconducting channels with coupling to a s-wave superconductor, interwire pairing and hopping gives rise to an effective two-channel p-wave system. We consider such a effective system with two kinds of p-wave pairing $\triangle_{\uparrow \uparrow}({\bf p})=\triangle_{\downarrow \downarrow}({\bf p})$ and $\triangle_{\uparrow \uparrow}({\bf p})=-\triangle_{\downarrow \downarrow}({\bf p})$. The spectrum is gapped for all $\mu>0$, gap closes at $\mu=0$.}
\label{setup}
\end{figure}

\section{Symmetry classification} \label{symmclass}
\subsection{Case I: $\triangle_{\uparrow\uparrow}=\triangle_{\downarrow\downarrow}$
}  \label{Sec1}
The first case we have studied is a BdG Hamiltonian (quadratic Hamiltonian describing gapped topological insulator and superconductor) with spinless p-wave superconductivity, such that the $ \triangle_{\uparrow\uparrow}=\triangle_{\downarrow\downarrow}$:
\begin{equation} \label{ham1}
\mathcal{H}_0(p) = \left(\frac{p^2}{2m}-\mu\right)\sigma_0\tau_z+\triangle_0 p\sigma_0\tau_x
\end{equation}
where, $\sigma$ and $\tau$ are Pauli matrices for the spin and the particle-hole sector, $\triangle_0$ and $\mu$ are effective parameters, depending on the underlying heterostructure. The Nambu spinor is $\begin{pmatrix}
c^{\dagger}_{p,\uparrow} & c_{-p,\uparrow} & c^{\dagger}_{p,\downarrow} & c_{-p,\downarrow}
\end{pmatrix}$, and $\triangle_0$ is the magnitude of the pairing $\triangle_{\uparrow\uparrow} (=\triangle_{\downarrow\downarrow})$. 

The symmetry class for Eq.~\ref{ham1} can be identified by studying the TR (time reversal), p-h (particle-hole) and chiral symmetries of the Hamiltonian. The symmetry operations are defined by: $\mathcal{T}\mathcal{H}(p)\mathcal{T}^{-1}=\mathcal{H}(-p)$, $\mathcal{P}\mathcal{H}(p)\mathcal{P}^{-1}=-\mathcal{H}(-p)$ and $\mathcal{S}\mathcal{H}(p)\mathcal{S}^{-1}=-\mathcal{H}(p)$. Using the TR operator $\mathcal{T}=\tau_z \mathcal{K}$, p-h operator $\mathcal{C}=\tau_x \mathcal{K}$ and chiral symmetry operator $\mathcal{S}=i\mathcal{T}.\mathcal{C}$, it can be seen that the Hamiltonian~\ref{ham1} lies in the BDI class, with $\mathcal{Z}$ topological invariant and integer number of edge states~\cite{schnyder}.

To get the explicit form of the edge states, Eq.~\ref{ham1} needs to be solved with the boundary condition $\psi(x=0)=0$, which gives four allowed values of $p$ (with $m,\triangle_0=1$):
\begin{equation}\label{pvalues}
p=\pm \sqrt{-1+\mu\pm\sqrt{1-2\mu}}
\end{equation}
Depending on whether $\mu>1/2$ or $\mu<1/2$, we get two different type of edge states. 
For $\mu>1/2$, the decaying modes are:
\begin{eqnarray} \label{p_mu1}
p_{1} &=& \sqrt{-1+\mu+\sqrt{1-2\mu}} \nonumber\\
p_2 &=& -\sqrt{-1+\mu-\sqrt{1-2\mu}}
\end{eqnarray}

Edge states with $E=0$:
\begin{align}\label{psi_mu1}
\psi_1 &= 2i \begin{pmatrix}
0\\
0\\
i\\
1\\
\end{pmatrix} e^{-\alpha x} \sin{\kappa x},&
\psi_2 &= 2i \begin{pmatrix}
i\\
1\\
0\\
0\\
\end{pmatrix} e^{-\alpha x} \sin{\kappa x},
\end{align}
with $\alpha=\rm{Im}(p_1)=\rm{Im}(p_2)$ and $\kappa=\rm{Re}(p_1)=-\rm{Re}(p_2)$. These edge states have both a decaying and oscillating nature, different from the usual purely decaying form.

For $0<\mu<1/2$, decaying modes: 
\begin{eqnarray} \label{p_mu2}
p_{1} &=& \sqrt{-1+\mu+\sqrt{1-2\mu}} \nonumber\\
p_3 &=& \sqrt{-1+\mu-\sqrt{1-2\mu}}
\end{eqnarray}
Purely decaying edge states with $E=0$:
\begin{align}\label{psi_mu2}
\psi_1 &= \begin{pmatrix}
0\\
0\\
i\\
1\\
\end{pmatrix} (e^{-\alpha_1 x}-e^{-\alpha_2 x}),&
\psi_2 &= \begin{pmatrix}
i\\
1\\
0\\
0\\
\end{pmatrix} (e^{-\alpha_1 x}-e^{-\alpha_2 x})
\end{align}
where, $\alpha_{1,2}=\rm{Im}(p_{1,2})$ and $\rm{Re}(p_1)=\rm{Re}(p_2)=0$. There are no zero energy edge states for $\mu<0$. In both the above cases, with $\mu>1/2$ and $0<\mu<1/2$, the edge states $\psi_1$ and $\psi_2$ are eigenstates of the p-h operator, and hence, are also Majorana Bound states (MBS). For this case, on adding perturbations like s-wave or magnetic field, the only non-trivial topological class that one can get is the BDI class. We enlist below all the possible combinations of magnetic fields that can preserve this symmetry class.

\subsubsection*{BDI class} \label{1_bd1}
This is a topologically non-trivial class (in 1D) in which the pure p-wave Hamiltonian in Eq.~\ref{ham1} belongs. Perturbations like the s-wave pairing $\triangle_1 \sigma_y\tau_y$ and the combinations of the two Zeeman fields ($\mathbf{B_1\hat{\sigma}\tau_z}$ and $\mathbf{B_2\hat{\sigma}\tau_0}$) can also generate the BDI class, with topological invariant $Z$.

\subsubsection*{\underline{$\mathbf{B_1.\hat{\sigma}\tau_z}$}}
In the absence of operator $\sigma$ in the p-wave Hamiltonian in Eq.~\ref{ham1}, any magnetic field can still be aligned along $\sigma_z$, and the Hamiltonian can be block diagonalized into two $2\times2$ irreducible blocks ($\sigma_z=\pm1$). However, due to the $\tau_z$ term, $\mathbf{B_1}$ does not get added as an overall constant in each block in this case. The chiral symmetry is still preserved in the blocks, and the system still remains in the same symmetry class BDI.
The allowed $p$-values will be of the same form as in Eq.~\ref{pvalues}, Eq.~\ref{p_mu1} and Eq.~\ref{p_mu2}, with different onsite potentials $\mu_{\uparrow,eff}=\mu_{\uparrow}-B_1$ for up-spin and $\mu_{\downarrow,eff}=\mu_{\uparrow}+B_1$ for down spin. 

\begin{figure*}
\begin{subfigure}
\centering
\includegraphics[width=6.5cm,height=5.5cm]{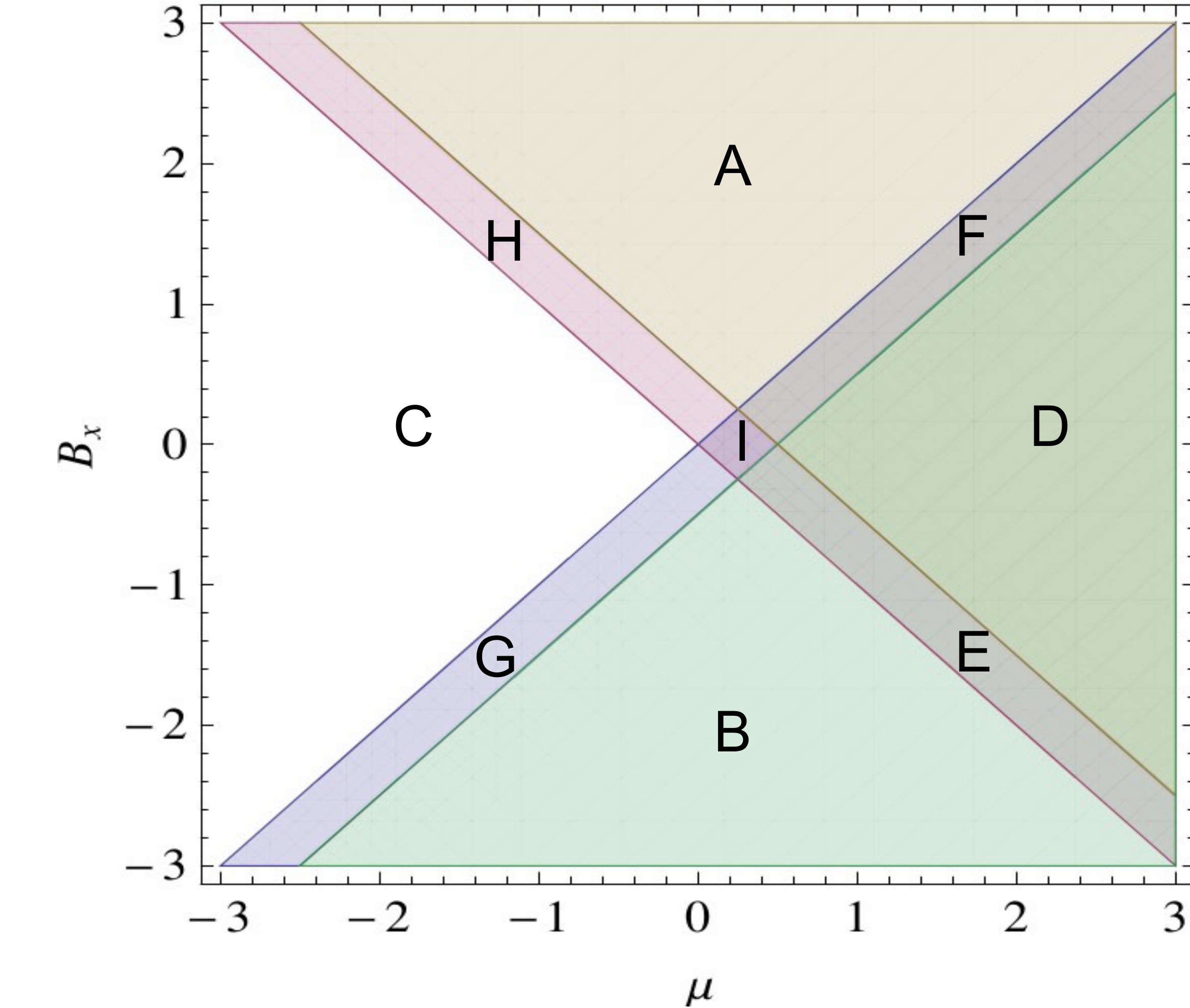}
\end{subfigure} 
\begin{subfigure}
\centering
\includegraphics[width=5.5cm,height=4.5cm]{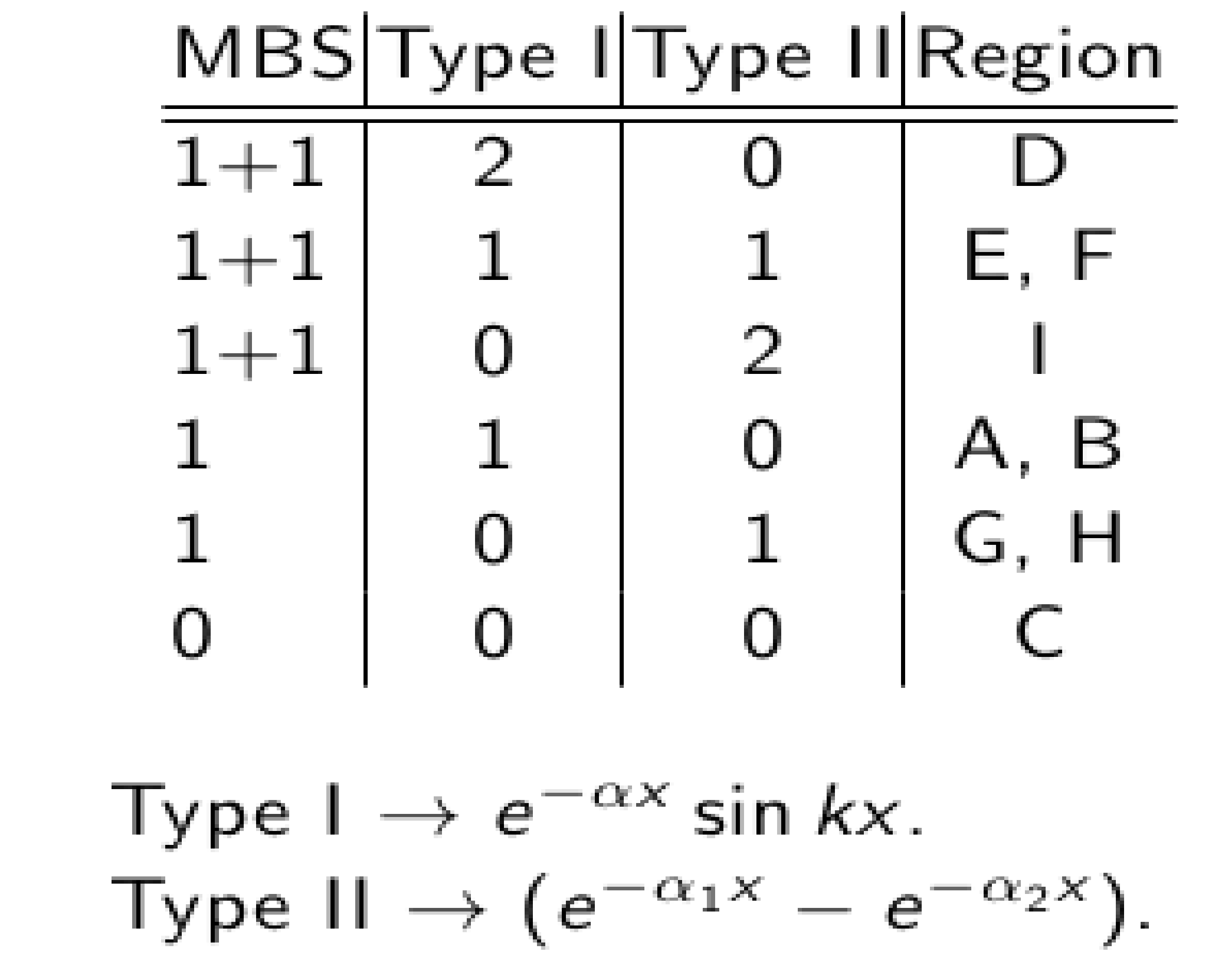}
\end{subfigure}
\caption{Left panel: Phase diagram in the $\mu-B_1$ space, showing the number of MBS. For the plot we have considered $\mathbf{B_1}$ along $\sigma_x$. However, as discussed, $\mathbf{B_1}$ along any other direction would still give the same phase diagram. Region C does not have any Majorana modes. All the other coloured regions have a Majorana singlet or a doublet, as indicated in the table on the right. Each of the modes can be of two types, purely decaying or damped oscillating, depending on which region of $\mu$ and $\mathbf{B}_1$ it is in. With the p-wave pairing $\triangle_{\uparrow\uparrow}=\triangle_{\downarrow\downarrow}$, and perturbation of the form $\mathbf{B}_1.\hat{\sigma}.\tau_z$, $\mathbf{B}_1$ along any direction will always give this phase diagram and all phases belong to the BDI class. }
\label{phase1}
\end{figure*}

The MBS have the same form as in Eq,~\ref{psi_mu1} and Eq,~\ref{psi_mu2}, now for particular ranges of values in the $\mu-B_1$ space, as shown in Fig.~\ref{phase1}.

\subsubsection*{\underline{$\triangle_1 \sigma_y\tau_y + B_1 \sigma_x\tau_z$ and $\triangle_1 \sigma_y\tau_y + B_1 \sigma_z\tau_z$}}
Here again the entire $4\times 4$ Hamiltonian, Eq.~\ref{ham1} along with the above perturbations, becomes irreducible and need to be considered in its entirety for symmetry classification. With $\triangle_1 \sigma_y\tau_y + B_1 \sigma_x\tau_z$, the TR operator is $\sigma_x\tau_z\mathcal{K}$, p-h operator is $\sigma_0\tau_x\mathcal{K}$ and chiral operator is $\sigma_x\tau_y$.  For $\triangle_1 \sigma_y\tau_y 
+ B_1 \sigma_z\tau_z$, the TR operator is $\sigma_z\tau_z\mathcal{K}$, p-h operator remains $\sigma_0\tau_x\mathcal{K}$ and chiral operator is $\sigma_z\tau_y$, giving $\mathcal{T}^2=1$, $\mathcal{C}^2=1$ and $\mathcal{S}=1$.

\subsubsection*{\underline{$B_1 \sigma_z\tau_z+B_2 \sigma_x\tau_0$ and $B_1 \sigma_z\tau_z+B_2 \sigma_y\tau_0$}}
When the two Zeeman fields $\mathbf{B}_1$ and $\mathbf{B}_2$ are simultaneously present in Eq.~\ref{ham1}, and $\mathbf{B}_1$ is fixed along $\hat{z}$, both $B_2 \sigma_x \tau_0$ and $B_2 \sigma_y \tau_0$ give the BDI class separately. For the first case the TR operator is $\tau_z\mathcal{K}$, and the p-h operator is
$\sigma_z\tau_x\mathcal{K}$.
For the second case the TR operator is $\sigma_z\tau_z\mathcal{K}$, and the p-h operator is
$\tau_x\mathcal{K}$. In both cases, the chiral operator is
$\tau_y$, giving $\mathcal{T}^2=1$, $\mathcal{C}^2=1$ and $\mathcal{S}=1$.

\begin{table*}[h]
\centering
\caption{Summary table for perturbations giving BDI class for Case I: $\triangle_0 p\sigma_0 \tau_x$}
\setlength{\tabcolsep}{13pt}
\begin{tabular}{l | c | c | c }
Perturbation & $\mathcal{T}$ & $\mathcal{C}$ & $\mathcal{S}$ \\
\hline \hline
$\mathbf{B}_1\hat{\sigma}\tau_z$ & 
\begin{minipage}{2.5cm}
$\tau_z\mathcal{K}$\\
$\mathcal{T}^2=1$
\end{minipage}
 & 
 \begin{minipage}{2.5cm}
$
\tau_x \mathcal{K}
$ \\
$\mathcal{C}^2=1$
\end{minipage}
& \begin{minipage}{2.5cm}
$
\tau_y
$ \\
$\mathcal{S}=1$
\end{minipage}
\\ 
\\
\hline
$\triangle_1 \sigma_y\tau_y+B_1\sigma_x\tau_z$  & 
\begin{minipage}{2cm}$
\sigma_x\tau_z\mathcal{K}
$\\
$\mathcal{T}^2=1$
\end{minipage}
 & 
 \begin{minipage}{2cm}
$ \tau_x\mathcal{K}
$ \\
$\mathcal{C}_{eff}=1$
\end{minipage}
& 
\begin{minipage}{2.5cm}
$
\sigma_x\tau_y
$ \\
$\mathcal{S}=1$
\end{minipage}
\\ 
\\
\hline
$\triangle_1 \sigma_y\tau_y+B_1\sigma_z\tau_z$ & 
\begin{minipage}{1.5cm}$
\sigma_z\tau_z \mathcal{K},
$
$\mathcal{T}^2=1$
\end{minipage}
 & 
 \begin{minipage}{1.5cm}$
\tau_x \mathcal{K},
$
$\mathcal{C}^2=1$
\end{minipage}
& 
\begin{minipage}{1.5cm}$
\sigma_z\tau_y,
$
$\mathcal{S}=1$
\end{minipage}
\\ 
\\
\hline
$B_1\sigma_z\tau_z + B_2\sigma_x\tau_0$& 
\begin{minipage}{2cm}
$\tau_z \mathcal{K}
$\\
$\mathcal{T}^2=1$
\end{minipage}
 & 
 \begin{minipage}{2cm}
$\sigma_z\tau_x \mathcal{K}
$\\
$\mathcal{C}^2=1$
\end{minipage}
& 
\begin{minipage}{2cm}
$\tau_y
$\\
$\mathcal{S}=1$
\end{minipage}
\\ 
\\
\hline
$B_1\sigma_z\tau_z + B_2\sigma_y\tau_0$& 
\begin{minipage}{2cm}
$\sigma_z\tau_z \mathcal{K}
$\\
$\mathcal{T}^2=1$
\end{minipage}
 & 
 \begin{minipage}{2cm}
$\tau_x \mathcal{K}
$\\
$\mathcal{C}^2=1$
\end{minipage}
& 
\begin{minipage}{2cm}
$\tau_y
$\\
$\mathcal{S}=1$
\end{minipage}
\\
\\
\hline
\end{tabular}

\label{table3}
\end{table*}

\subsection{Case II: $\triangle_{\uparrow\uparrow}=-\triangle_{\downarrow\downarrow}$
} \label{Sec2}

The second case we have studied is a BdG Hamiltonian with spinless p-wave superconductivity, such that the $ \triangle_{\uparrow\uparrow}=-\triangle_{\downarrow\downarrow}$:
\begin{equation} \label{ham2}
\mathcal{H}_0(p) = \left(\frac{p^2}{2m}-\mu\right)\sigma_0\tau_z+\triangle_0 p\sigma_z\tau_x
\end{equation}
where, $\sigma$ and $\tau$ are Pauli matrices for the spin and the particle-hole sector, and $\triangle_0$ and $\mu$ are the effective parameters. The Nambu spinor is again, $\begin{pmatrix}
c^{\dagger}_{p,\uparrow} & c_{-p,\uparrow} & c^{\dagger}_{p,\downarrow} & c_{-p,\downarrow}
\end{pmatrix}$, and $\triangle_0$ is the magnitude of the pairing $\triangle_{\uparrow\uparrow} (=|\triangle_{\downarrow\downarrow}|)$.

The symmetry class can be identified by studying the TR (time reversal), p-h (particle-hole) and chiral symmetries of the Hamiltonian. The symmetry operations are defined by: $\mathcal{T}\mathcal{H}(p)\mathcal{T}^{-1}=\mathcal{H}(-p)$, $\mathcal{P}\mathcal{H}(p)\mathcal{P}^{-1}=-\mathcal{H}(-p)$ and $\mathcal{S}\mathcal{H}(p)\mathcal{S}^{-1}=-\mathcal{H}(p)$. However, in this case there is more than one possible TR and p-h operator each: $\mathcal{T}=\sigma_x\mathcal{K}$, $\sigma_y\mathcal{K}$ and $\tau_z \mathcal{K}$, and, $\mathcal{C}=\tau_x \mathcal{K}$, $\sigma_z\tau_x \mathcal{K}$, $\sigma_y\tau_y \mathcal{K}$ and $\sigma_x\tau_y \mathcal{K}$. By block diagonalizing the Hamiltonian in Eq.~\ref{ham2} into $\sigma_z=\pm1$ blocks, the effective TR operator is $\tau_z\mathcal{K}$ and p-h operator is $\tau_x\mathcal{K}$. It can be seen that the Hamiltonian lies in the BDI class ($\mathcal{T}^2=1$, $\mathcal{C}^2=1$ and $\mathcal{S}=1$) , with $\mathcal{Z}$ topological invariant and integer number of edge states, see Ref.~\cite{tenfold} and Appendix~\ref{redef_symm}.

The allowed $p$ values have the same form as Eq.~\ref{pvalues}. Here again, we have two MBS $\psi_1$ and $\psi_2$, Eq.~\ref{psi_mu1} (for $\mu>1/2$) and Eq.~\ref{psi_mu2} (for $0<\mu<1/2$). Only the structure of the eigenvectors of the MBS is different:
\begin{align} \label{MBS_2}
\begin{pmatrix}
0\\
0\\
i\\
1\\
\end{pmatrix}, &\ \text{and},\ \begin{pmatrix}
-i\\
1\\
0\\
0\\
\end{pmatrix} 
\end{align}
The phase diagram for the MBS in this type of p-wave SC is also the same as in Fig.~\ref{phase1}.
Below, we list the possible non-trivial topological classes that can arise from certain combinations of s-wave and Zeeman terms. For the trivial classes, refer to Aec.~\ref{appendix}.

\subsubsection{AIII class} \label{2_a3}
AIII symmetry class is a topologically non-trivial class, with $Z$ invariant. The symmetry conditions are : $\mathcal{T}=0$, $\mathcal{C}=0$ and $\mathcal{S}=1$. Spinless p-wave SC of the particular type in Eq.~\ref{ham1} can access this symmetry class in the presence of certain perturbations like the s-wave $\mathbf{\triangle_1 \sigma_y\tau_y}$, and combinations of the s-wave and Zeeman term along $\tau_0$.

\subsubsection*{\underline{$\mathbf{\triangle_1 \sigma_y\tau_y}$}}
In the presence of a s-wave term in Eq.~\ref{ham2}:
\begin{equation} \label{ham2_2}
\mathcal{H}(p)=\mathcal{H}_0(p)+\triangle_1 \sigma_y\tau_y
\end{equation}
TR operator, satisfying the condition  $\mathcal{T}\mathcal{H}(p)\mathcal{T}^{-1}=\mathcal{H}(-p)$, is $\sigma_y\mathcal{K}$. 
There are two possible p-h operator for Eq.~\ref{ham2_2}, $\tau_x\mathcal{K}$ and $\sigma_x\tau_y\mathcal{K}$. On block diagonalization into $\sigma_z=\pm1$ irreducible blocks, we get,
\begin{equation} \label{ham2_3}
H_{1,2}=-\left(\frac{p^2}{2m}-\mu\right)\tau_z-\triangle_0 p\tau_x\pm\triangle_1\tau_x
\end{equation}
However, each block in Eq.~\ref{ham2_3}, no longer has TR or p-h symmetry, since neither the two p-h operators nor the TR operator is a symmetry. But a chiral operator still exists, $\mathcal{S}=\tau_y$, which gives, $\mathcal{S}^2=1$. This class of Hamiltonian falls in the topologically non-trivial AIII symmetry class (in $d=1$), with $Z$ invariant.
Edge states in the basis $\begin{pmatrix}
   c^{\dagger}_{p,\downarrow}& c_{-p,\uparrow} & c^{\dagger}_{p,\uparrow}  & c_{-p,\downarrow}
\end{pmatrix}$ and $E=0$:
\begin{align}\label{psi_aiii}
\psi_1 &= 2i \begin{pmatrix}
0\\
0\\
i\\
1\\
\end{pmatrix} e^{- x} \sin{(\kappa + i \delta \kappa) x}, \nonumber \\
\psi_2 &= 2i \begin{pmatrix}
i\\
1\\
0\\
0\\
\end{pmatrix} e^{- x} \sin{(\kappa + i \delta \kappa) x},
\end{align}
with $\kappa=\text{Re}(\sqrt{-1 + 2 i \triangle_1 + 2 \mu})$ and  $\delta \kappa=\text{Im}(\sqrt{-1 + 2 i \triangle_1 + 2 \mu})$. These edge states have both a decaying and oscillating nature (similar to Type I in the BDI case), again, different from the usual purely decaying form.

\subsubsection*{\underline{$\triangle_1 \sigma_y\tau_y + B_2\sigma_x\tau_0$}}
With the above perturbation, when both the s-wave term and the zeeman term  $B_2\sigma_x\tau_0$ are present in Eq.~\ref{ham2}, we again get back the AIII class. $\mathcal{T}$ and $\mathcal{C}$ is zero, but $\mathcal{S}$ in each block is $\sigma_y\tau_x$, giving $\mathcal{S}=1$.

\begin{table*}[h]
\centering
\caption{Summary table for perturbations giving AIII class for Case II ($\triangle_0 p\sigma_z \tau_x$)}
\setlength{\tabcolsep}{13pt}
\begin{tabular}{l | c | c | c }
Perturbation & $\mathcal{T}$ & $\mathcal{C}$ & $\mathcal{S}$ \\
\hline \hline
$\triangle_1 \sigma_y\tau_y$ & 
\begin{minipage}{2.5cm}
$\sigma_y\mathcal{K}$\\
$\mathcal{T}_{eff}=0$
\end{minipage}
 & 
 \begin{minipage}{2.5cm}
$\tau_x\mathcal{K},\
\sigma_x\tau_y\mathcal{K}$\\
$\mathcal{C}_{eff}=0$
\end{minipage}
&
\begin{minipage}{2.5cm}
$\tau_y$\\
$\mathcal{S}=1$
\end{minipage}
\\ 
\\
\hline
$\triangle_1\sigma_y\tau_y +B_2\sigma_x\tau_0$  & 
0
 & 
0
& 
\begin{minipage}{2.5cm}
$\sigma_y\tau_x$\\
$\mathcal{S}=1$
\end{minipage}
\\ 
\\
\hline
\end{tabular}
\label{table5}
\end{table*}


\begin{figure*}[h]
\centering
\text{Phase diagram for edge states in AIII class}\par\medskip
\begin{tabular}{c}
\includegraphics[width=5.5cm, height=3.5cm]{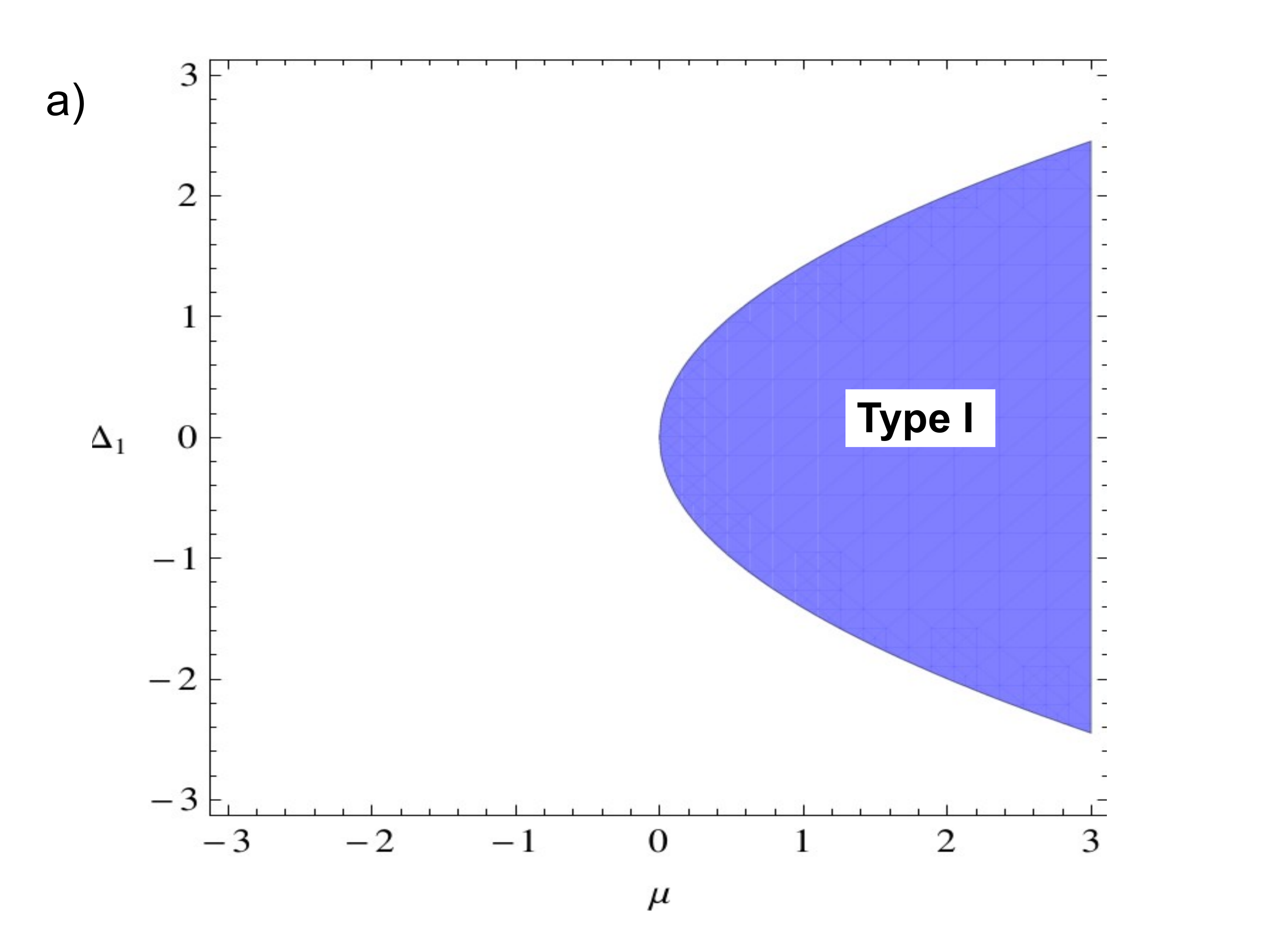}
\includegraphics[width=5.5cm, height=3.5cm]{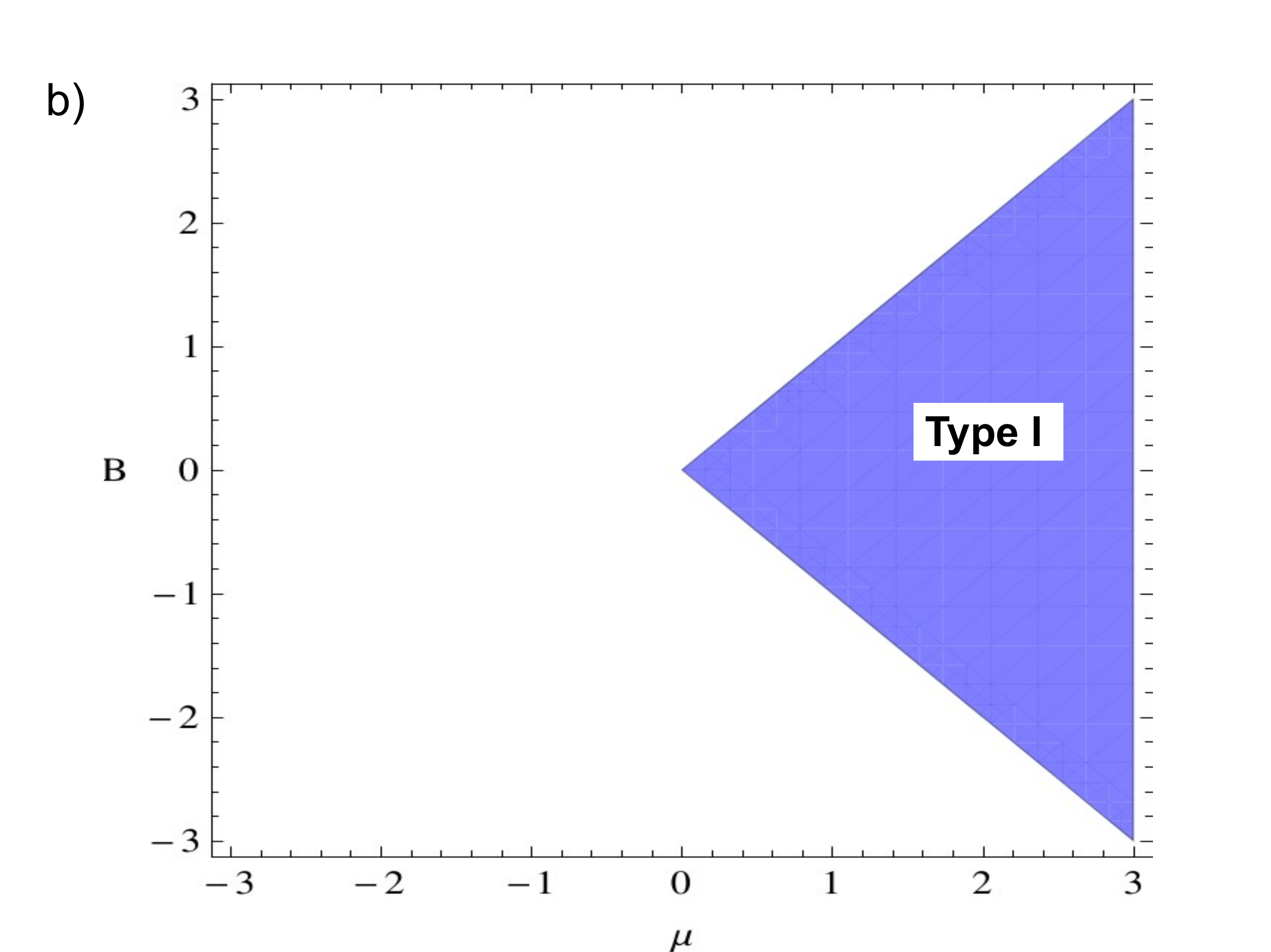} \\
\includegraphics[width=6cm, height=4.5cm]{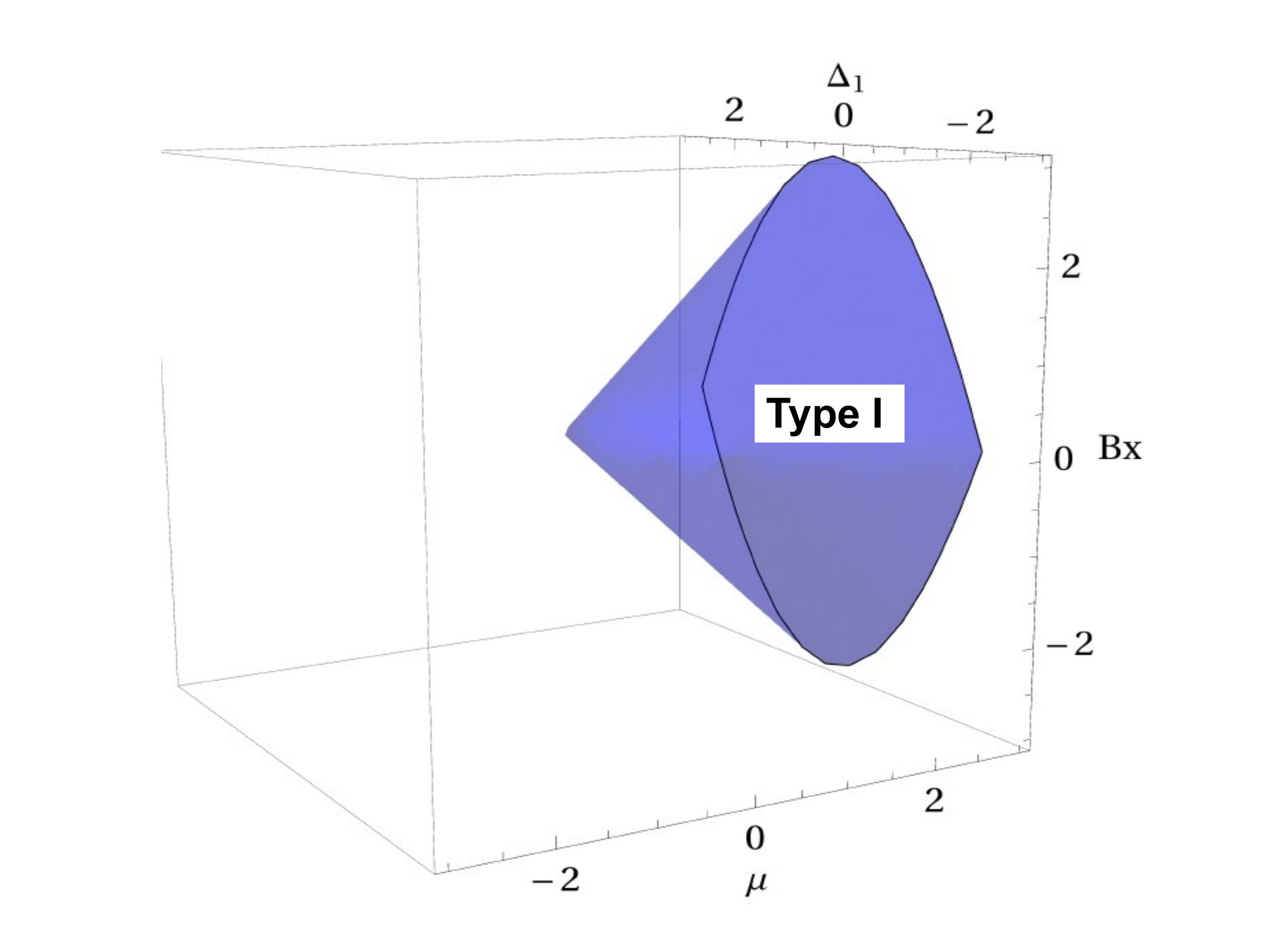}
\end{tabular}
\caption{Phase diagram for MBS in the presence of s-wave perturbations in Eq.~\ref{ham2}. (Top Left) Region plot showing the regions in $\mu-\triangle_1$ space over which edge states of Type I exist in the presence of s-wave. Top Right: Region plot in $\mu-B$ space over which again edge states of Type I exist. Bottom: Three dimensional region plot in $\mu-\triangle_1-B$ space, showing edge states exists only for regions above $\mu>0$, all belonging to the Type I category.}
\label{phase_swave}
\end{figure*}


\subsubsection{BDI class} \label{2_bd1}
As discussed earlier in Sec.~\ref{1_bd1}, the topologically non-trivial BDI class can appear in 1D p-wave SC. However, with the particular type in Eq.~\ref{ham2}, an s-wave pairing term cannot induce such a transition. It is necessary for the Zeeman terms, along $\tau_0$ and $\tau_z$, to be present. The symmetry conditions are $\mathcal{T}^2=1$, $\mathcal{C}^2=1$ and $\mathcal{S}=1$.

\subsubsection*{\underline{$\mathbf{B_2.\hat{\sigma}\tau_0}$}}
\begin{itemize}
\item $B_2\sigma_x\tau_0$: Possible TR operators are $\sigma_x\mathcal{K}$ and $\tau_z\mathcal{K}$, with the effective $\mathcal{T}$ in each block being $\tau_z\mathcal{K}$. Similarly, possible p-h operators are $\sigma_z\tau_x\mathcal{K}$ and $\sigma_y\tau_y\mathcal{K}$, with the effective $\mathcal{C}$ being $\tau_x\mathcal{K}$. Chiral operator $\mathcal{S}$ is $\tau_y$.

\item $B_2\sigma_y\tau_0$: Here again the effective $\mathcal{T}$ is $\tau_z\mathcal{K}$. The possible p-h operators are $\tau_x\mathcal{K}$ and $\sigma_y\tau_y\mathcal{K}$, with the effective $\mathcal{C}$ being $\tau_x\mathcal{K}$. Chiral operator $\mathcal{S}$ is $\tau_y$.
\end{itemize}

\subsubsection*{\underline{$B_1\sigma_z\tau_z$}}
The TR operator $\mathcal{T}$ is $\tau_z\mathcal{K}$. The possible p-h operators are $\tau_x\mathcal{K}$ and $\sigma_z\tau_x\mathcal{K}$, with the effective $\mathcal{C}$ being $\tau_x\mathcal{K}$. Chiral operator $\mathcal{S}$ is $\tau_y$.


\begin{table*}[h]
\centering
\caption{Summary table for perturbations giving BDI class for Case II ($\triangle_0 p\sigma_z \tau_x$)}
\setlength{\tabcolsep}{13pt}
\begin{tabular}{l | c | c | c }
Perturbation & $\mathcal{T}$ & $\mathcal{C}$ & $\mathcal{S}$ \\
\hline \hline
$B_2\sigma_x\tau_0$ & 
\begin{minipage}{2.5cm}
$\sigma_x\mathcal{K},\
\tau_z\mathcal{K}
$\\
$\mathcal{T}_{eff}=\tau_z\mathcal{K}$\\
$\mathcal{T}^2=1$
\end{minipage}
 & 
 \begin{minipage}{2.5cm}
$\sigma_y\tau_y\mathcal{K},\
\sigma_z\tau_x\mathcal{K}$\\
$\mathcal{C}_{eff}=\tau_x\mathcal{K}$\\
$\mathcal{C}^2=1$
\end{minipage}
&
\begin{minipage}{2.5cm}
$\tau_y$\\
$\mathcal{S}=1$
\end{minipage}
\\ 
\\
\hline
$B_2\sigma_y\tau_0$  & 
\begin{minipage}{2cm}
$\mathcal{T}_{eff}=\tau_z\mathcal{K}$\\
$\mathcal{T}^2=1$
\end{minipage}
 & 
\begin{minipage}{2.5cm}
$\sigma_y\tau_y\mathcal{K},\
\tau_x\mathcal{K}$\\
$\mathcal{C}_{eff}=\tau_x\mathcal{K}$\\
$\mathcal{C}^2=1$
\end{minipage}
& 
\begin{minipage}{2.5cm}
$\sigma_y\tau_x$\\
$\mathcal{S}=1$
\end{minipage}
\\ 
\\
\hline
$B_1\sigma_z\tau_z$  & 
\begin{minipage}{2cm}
$\tau_z\mathcal{K}$\\
$\mathcal{T}^2=1$
\end{minipage}
 & 
\begin{minipage}{2.5cm}
$\sigma_z\tau_x\mathcal{K},\
\tau_x\mathcal{K}$\\
$\mathcal{C}_{eff}=\tau_x\mathcal{K}$\\
$\mathcal{C}^2=1$
\end{minipage}
& 
\begin{minipage}{2.5cm}
$\sigma_y\tau_x$\\
$\mathcal{S}=1$
\end{minipage}
\\ 
\\
\hline
\end{tabular}
\label{table6}
\end{table*}


\subsubsection{D class} \label{2_d}
The D symmetry class is another topologically non-trivial class in 1D, with $\mathcal{T}=0$, $\mathcal{C}^2=1$ and $\mathcal{S}=0$, with the invariant being $Z_2$. The unperturbed p-wave Hamiltonian in Eq.~\ref{ham2} can access this class in the presence of s-wave and Zeeman terms $B_1\sigma_z\tau_z$ and  $B_2\sigma_y\tau_0$ , and also with combinations of both the Zeeman term $\mathbf{B}_1$ and $\mathbf{B}_2$.

\subsubsection*{\underline{$\triangle_1 \sigma_y\tau_y + B_2\sigma_y\tau_0$ and $\triangle_1 \sigma_y\tau_y + B_1\sigma_z\tau_z$}}
With each of the above perturbations, the p-h operator $\mathcal{C}$ is $\tau_x\mathcal{K}$. However, since the chiral symmetry is absent the edge states do not appear at $E=0$ and is not captured in the present MBS calculation.

\subsubsection*{\underline{$B_1(\sigma_x +\sigma_z) \tau_z+B_2\sigma_y\tau_0$}}
Here again we consider the combination of two Zeeman fields in Eq.~\ref{ham2}, $\mathbf{B}_1 \tau_z$ in the $x$-$z$ plane and $\mathbf{B}_2\tau_0$ along $\sigma_y$. Here again, the p-h operator $\mathcal{C}$ is $\tau_x\mathcal{K}$.


\begin{table*}[h]
\centering
\caption{Summary table for perturbations giving D class for Case II ($\triangle_0 p\sigma_z \tau_x$)}
\setlength{\tabcolsep}{13pt}
\begin{tabular}{l | c | c | c }
Perturbation & $\mathcal{T}$ & $\mathcal{C}$ & $\mathcal{S}$ \\
\hline \hline
$\triangle_1\sigma_y\tau_y+B_2\sigma_y\tau_0$\\
$\triangle_1\sigma_y\tau_y+B_1\sigma_z\tau_z$ & 
0
 & 
 \begin{minipage}{1.5cm}
$\tau_x\mathcal{K}$\\
$\mathcal{C}^2=1$
\end{minipage}
&
0
\\ 
\\
\hline
$B_1(\sigma_x+\sigma_z)\tau_z+B_2\sigma_y\tau_0$  & 
0
 & 
\begin{minipage}{1.5cm}
$
\tau_x\mathcal{K}$\\
$\mathcal{C}^2=1$
\end{minipage}
& 
0
\\ 
\\
\hline
\end{tabular}
\label{table7}
\end{table*}


\section{Subgap states across coupled p-wave platforms} \label{JJ}

\subsection{BDI class} \label{subgap_BDI}
	
We consider a junction between two spinless p-wave superconductors in 1D. On both sides of the junction we have p-wave superconductor, belonging to BDI class, but there is a phase difference between the order parameters. We consider the junction barrier to be located at $x=0$, having the form $U(x) = U_0 \delta(x)$~\cite{KSG}. One can also use the scattering matrix approach in Ref.~\cite{beenakker, beri}, which is essentially the same as the procedure we use here, but with a $\delta$-function potential. 

\begin{figure}[H]
	\centering
	\includegraphics[width=0.8\columnwidth]{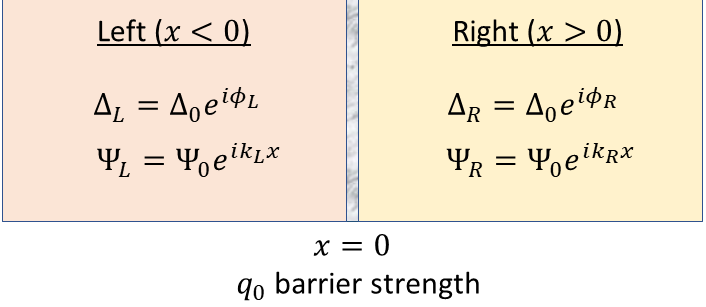}
	\caption{For our analysis, our junction is between two topological superconductors with p-wave order parameters $\triangle_R$ and $\triangle_L$ respectively. The simplest combination is between two platforms belonging to BDI class, with a phase difference between the order parameters on the two sides. On the subsequent analyses, we consider effect of different perturbations on each side, like an additional s-wave term or a magnetic field, which changes the topological class on each side. An effective barrier strength is considered, given by $q_0$.}
\end{figure}

The Hamiltonian on each side of the junction is:
\be \label{ham1}
\mathcal{H_\nu}=\left(\frac{p^2}{2m}-\mu\right)\sigma_0\tau_z
+ \triangle_0 p \sigma_0 (\cos{\phi_\nu}  \tau_x + i \sin{\phi_\nu}  \tau_y) ,
\ee
where $\nu\equiv\left[R,L\right]$. Here, we show the calculation for the first kind of p-wave, given as Case I in Sec.~\ref{Sec1}. However, one can do an exact same calculation for the second kind (given in Case II, Sec~\ref{Sec2}) and get the same form of the subgap states, with the eigenvectors in Eq.~\ref{solutions} having the structure given in Eq.~\ref{MBS_2}. 

Since there is no $\sigma_{x,y}$ term in the Hamiltonian, we consider a single spin species and the trial wave function on each side of the junction is:
\beq \label{trial_wf}
\psi_\nu &=& \psi_0 e^{i k_\nu x} \nonumber \\
&=& \begin{pmatrix}
        u_\nu(k_\nu)\\
        v_\nu(k_\nu)\\
       \end{pmatrix} e^{i k_\nu x},
\eeq
where, $k_\nu = k_\nu^r+i k_\nu^I$ in general. The real part becomes $e^{i k_\nu^r x}$, giving a plane wave behaviour and the imaginary part becomes $e^{- k_\nu^I x}$, giving the decaying part. In order to get a decaying solution as $x\rightarrow\infty$, the imaginary part should be positive on the right side of the junction ($x>0$) and negative on the left hand side of the junction ($x<0$). Thus, when we solve the eigenvalue equation for Eq.~\ref{ham1} on each side of the junction, we choose only those values of $k_\nu$  for which $k_R^I>0$ and $k_L^I<0$. (We will have 4 possible solutions for $k_\nu$ on each side, this criterion allows us to pick two among them, on each side.)

Using Eq.~\ref{ham1} and Eq.~\ref{trial_wf}, the eigenvalue equation can be obtained, which gives the possible expressions for $k_\nu$ as a function of energy $E$. A general wave function on each side of the junction will be a linear combination of the two possible $k_\nu$ modes~\cite{KSG}.

\beq \label{solutions}
\Psi_R &=& A_1\begin{pmatrix}
        u(k_{R,1})\\
        v(k_{R,1})\\
       \end{pmatrix} e^{i k_{R,1} x}+B_1\begin{pmatrix}
        u(k_{R,2})\\
        v(k_{R,2})\\
       \end{pmatrix} e^{i k_{R,2} x} \nonumber \\
\Psi_L &=& A_2\begin{pmatrix}
        u(k_{L,1})\\
        v(k_{L,1})\\
       \end{pmatrix} e^{i k_{L,1} x}+B_2\begin{pmatrix}
        u(k_{L,2})\\
        v(k_{L,2})\\
       \end{pmatrix} e^{i k_{L,2} x},       
\eeq
where $k_R$ has a positive imaginary part and $k_L$ has negative imaginary part. Using Eq.~\ref{solutions} in the following boundary conditions:
\beq \label{boundarycond}
\Psi_R(x=0)&=&\Psi_L(x=0) \nonumber\\
\frac{\partial\Psi_R}{\partial x}\bigg|_{x=0} -\frac{\partial\Psi_L}{\partial x}\bigg|_{x=0}&=& q_0 \Psi_R(x=0),
\eeq
(where, $q_0 \propto U_0$), we get the consistency condition that the determinant from Eq.~\ref{boundarycond} should be equal to zero. This gives an equation of the form: $f(E,\mu,\triangle_0,q_0,\phi)=0$, from where we get the solutions for $E$ as a function of $\phi$. ($\phi=\phi_R-\phi_L$)

\begin{figure}[H]
\centering
\textbf{Subgap states across BDI class}\par\medskip
\includegraphics[width=0.48\linewidth]{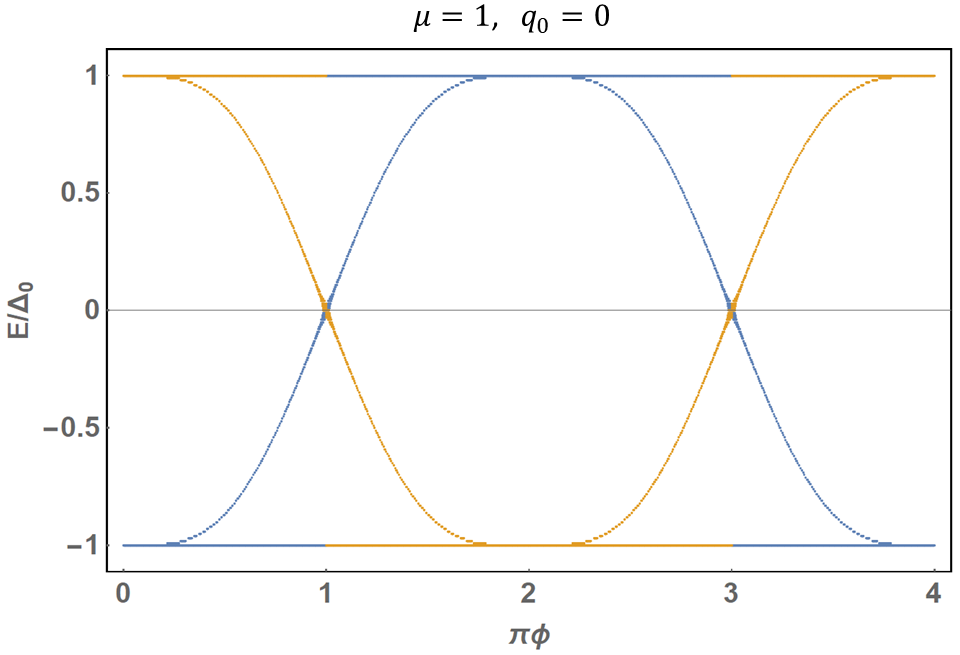}
 \includegraphics[width=0.48\linewidth]{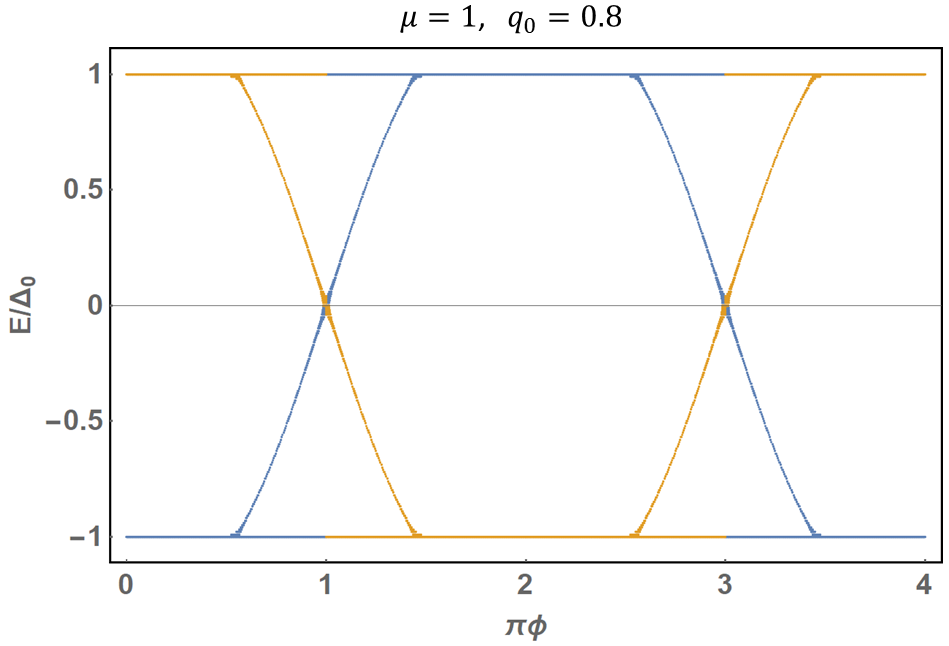} \\
\vspace{0.1cm}

\includegraphics[width=0.49\linewidth]{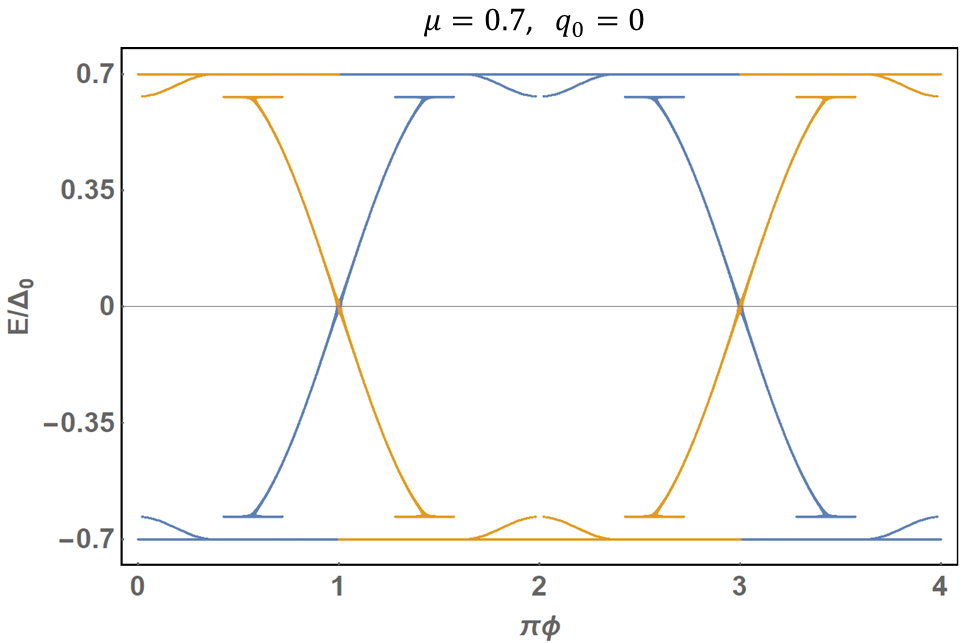}
 \includegraphics[width=0.48\linewidth]{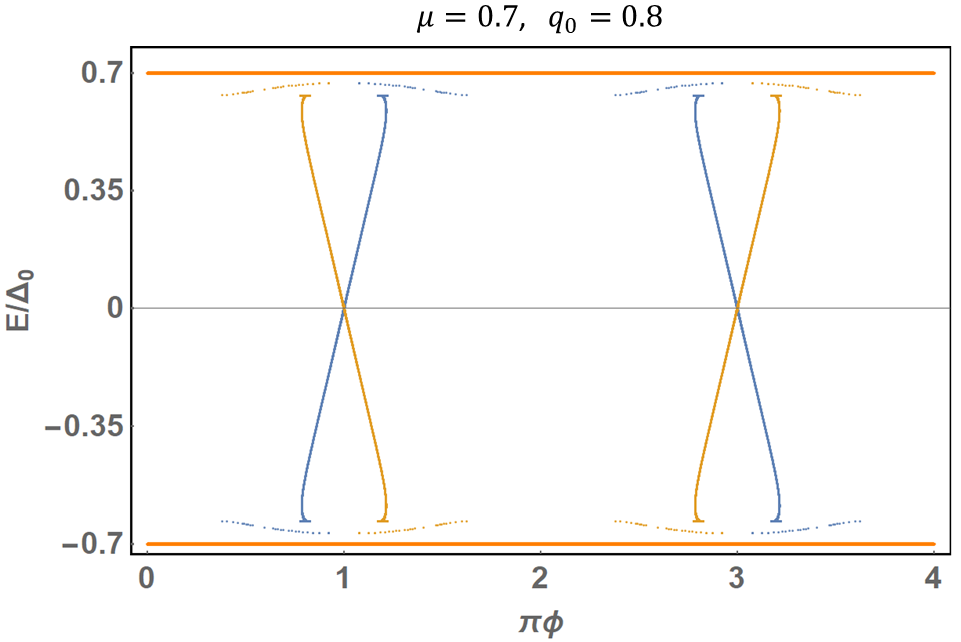} \\
\vspace{0.1cm}

 \includegraphics[width=0.48\linewidth]{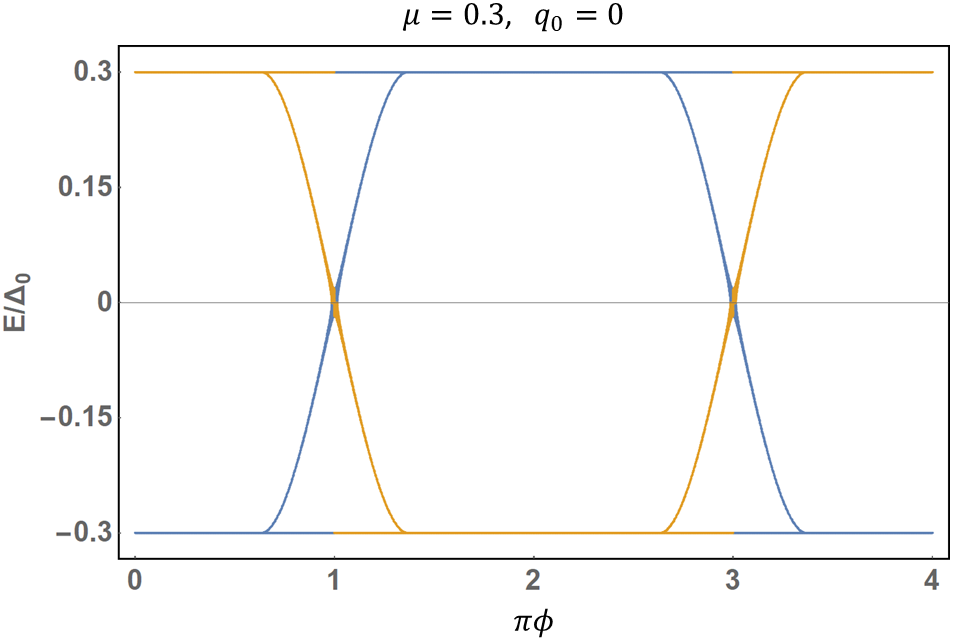}
 \includegraphics[width=0.48\linewidth]{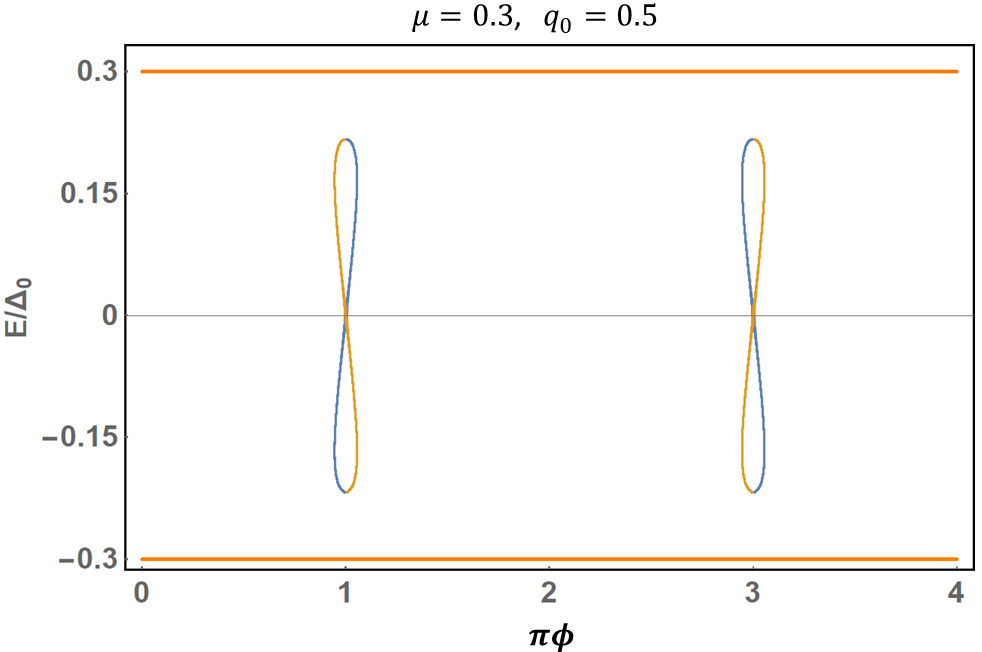}
  \caption{Subgap states across junctions of BDI class, with $\triangle_0=1$. Top row: for $\mu=1$, middle row: for $\mu=0.7$, bottom row: for $\mu=0.3$. Plots in the left column are for $q_0=0$ and right column for non-zero barrier. 
For $\mu=1$ and zero barrier, the subgap states are $\sim \cos{\phi/2}$. When $q_0 \neq 0$, the subgap states are pushed into the bulk, with the upper and bottom part of the curves being chopped off. However, they are still $4\pi$ periodic. When $\mu<1$, `wing' like additional states appear close to the bulk bands. With stronger barrier, the additional states get erased for some values of $\phi$ and appear for other. For non-zero $q_0$, the orange bands denote the bulk bands separately for $\mu=0.7$ and $\mu=0.3$. When $\mu=0.7$, the subgap states do not merge with the bulk bands, even when barrier is zero.  When $\mu<0.5$, the zero barrier subgap states resemble the ones with $\mu=1$ in the strong barrier regime. With stronger barrier, more subgap states get erased before reaching the bulk in $\mu<1/2$ case.}
\label{mu_bdI}
\end{figure}

In Fig.~\ref{mu_bdI}, we have calculated $E$ vs. $\phi$ when both Hamiltonians belong to BDI.  BDI class has integer number of edge modes, which hybridize across the junction to give a solution as a function of $\phi$. Since $\mu$ in this model is an effective parameter, we do not restrict to the limit when $\mu \rightarrow \infty$. This illuminates more finer features about the subgap states. We look in both limits of the barrier strength, $q_0=0$ and $q_0$ is high. We see that some features for the strong barrier case when $\mu>\triangle_0^2/2$ is reciprocated in the weak barrier case when $\mu<\triangle_0^2/2$ (i.e, $\mu<1/2$, when $\triangle_0=1$). 

Without loss of generality, we put $\triangle_0=1$ and measure all quantities with respect to $\triangle_0$. In these units, when $\mu=1$, the solutions for the midgap states are shown in Fig.~\ref{mu_bdI} top panel. If we stick to one branch, it can be seen that energy is $4\pi$ periodic, and is $\sim\cos{\phi/2}$. The bulk bands are at $E=\pm \mu$. When barrier is non-zero, the subgap states are pushed into the bulk bands and the top and bottom parts of the curve get cut off. 

When $1/2<\mu<1$, the subgap states are plotted in the middle panel of Fig.~\ref{mu_bdI}. They show a similar $4\pi$ periodicity as seen in the previous case, which becomes more apparent when we analytically solve for the solutions putting $q_0=0$, in Sec.~\ref{analytical_subgap}. Also, we get additional solutions at $E \ge |\sqrt{-1+2\mu}|$. It is interesting to note that for these $E$'s, the form of the subgap states Eq.~\ref{solutions} changes. 
That is, one $k$, on each side of the junction, is no longer decaying (has the form $\sim e^{\alpha x}$ for $x>0$), and does not give a physical solution. The other remaining $k$, that gives a decaying solution, is the only physically valid subgap solution. This gives a different set of solutions for the midgap states in the regime $E \ge |\sqrt{-1+2\mu}|$ and $E\le \mu$. This is what we see in the figure, where for $\mu=0.7$, we see additional states as a function of $\phi$ (additional `wings') close to the bulk bands. We also see the subgap states for $E < |\sqrt{-1+2\mu}|$ end before reaching the bulk bands, i.e, there is a gap between these subgap states and the `wing' like states. The energy scale at which this happens is at $E_{c}= |\sqrt{-1+2\mu}|$, after which there is a sudden change in the functional form of the wave function and also the determinant  $f(E,\mu,\triangle_0,q_0,\phi)$. For $E> E_{c}$, the minima of the determinant suddenly becomes at $\phi=0$, $2\pi$ and $4\pi$, instead of $\pi$ and $3\pi$. Also, as the barrier strength increases, the nature of these `wing' like states changes. This can be understood from the boundary condition in this region, which effectively gives: $\frac{A_1}{A_2} = (1-q_0) \frac{u(k_{L,1})k_L}{u(k_{R,1}) k_R} $. $u(k_{L,1})$ and $u(k_{R,1})$ are also functions of $\phi_L$ and $\phi_R$, respectively. Depending on $\phi_R$ and $\phi_L$, a strong barrier can erase the additional states at some $\phi$, and generate them at other $\phi$'s (see, Fig.~\ref{mu_bdI} middle row). 

For $\mu<1/2$, the nature of the subgap states change considerably, see Fig.~\ref{mu_bdI} bottom row. Firstly, in the transparent barrier regime, the subgap states merge with the bulk bands ($E=\mu$) earlier, similar to strong barrier regime when $\mu=1$ . Again, we can derive the solutions analytically in the limit $q_0=0$ to see that the periodicity is still $4\pi$. Additionally, the nature of the states close to the bulk band do not change drastically. With increasing barrier strength, there is a smooth change of the determinant function and the two minimas merge to one as we move to higher energies. After the energy where it reaches a single minima (as shown by the top and bottom turning points of the lobes), the wavefunctions are no longer valid subgap states. Also with higher barrier strengths, the we get less and less subgap states, shown by the shrinking lobes at high $q_0$.

\subsection{AIII class}
The Hamiltonian on each side of the junction is now of the form Eq.~\ref{ham2_2}. Since in the AIII class there is only one type of zero energy edge states and for all values of $\mu>0$ (shown in Fig.~\ref{phase_swave}), we can linearize around the Fermi points and use the approach described in Ref.~\cite{beenakker, beri} to derive the Andreev reflection matrices across superconducting junctions. We rewrite the Hamiltonian in Eq.~\ref{ham2_2} in the basis $\begin{pmatrix}
c^{\dagger}_{p,\uparrow} & c_{p,\downarrow} & c^{\dagger}_{p,\downarrow} & -c^\dagger_{-p,\uparrow}
\end{pmatrix}$. The s-wave term takes the form $ \triangle_1 \sigma_0 \tau_x$. The Hamiltonian on each side is of the form:
\begin{equation}
\mathcal{H}_{AIII} = \begin{pmatrix}
h_e & \triangle \\
\triangle^{\dagger} & h_h\\
\end{pmatrix}
\end{equation}
where, $h_e$ and $h_h$ are the single-particle Hamiltonians for particle and hole sectors, respectively, and $\triangle$ is the superconducting pair potential. The barrier strength can be tuned to take into account a delta function potential.
Between two BDI materials (and also between two AIII materials), when there is two different $\mu$ on the two sides of the junction, there are no subgap states. Also, when there is BDI and AIII on each side of a junction, there are no subgap states. 

\section{Conclusion}
The two cases of the p-wave pairing in Sec.~\ref{Sec1} and Sec.~\ref{Sec2} have very distinct features in the presence of perturbations, even though both the unperturbed Hamiltonians start from the topologically non-trivial BDI class. Firstly, in the presence of the s-wave pairing term alone, the first type of p-wave Hamiltonian ($\triangle_{\uparrow\uparrow}=\triangle_{\downarrow\downarrow}$) becomes topologically trivial (AI) whereas the second type ($\triangle_{\uparrow\uparrow}=-\triangle_{\downarrow\downarrow}$) shifts to another non-trivial class, AIII. Secondly, the first type of p-wave can still remain in the BDI class in the presence of certain combinations of the Zeeman fields and the s-wave term. Whereas, the second type of p-wave cannot remain in the BDI class when the s-wave perturbation is present. This is true for also the AI class, which the second type of p-wave cannot access in the presence of s-wave pairing, but the first type can.

We have also shown that the MBS in each of the two unperturbed p-wave cases have the same spatial dependence, with the emergence of two types of Majorana modes depending on the value of $\mu$. For $\mu>1/2$ we get damped oscillating modes and for $0<\mu<1/2$ the MBS are purely decaying. The only difference between them is that the eigenvectors have a different structure for the two cases. This nature is also reflected in the presence of a Zeeman field, where both the p-wave cases have the same phase diagram, Fig.~\ref{phase1}.

The presence of an s-wave term in each of the cases changes the nature of edge states. For cases with s-wave which lie in the non-trivial topological classes with chiral symmetry (like the AIII class), zero energy MBS still appear for $\mu>0$, but they are only of the damped oscillating form, shown in Fig.~\ref{phase_swave}. This is another unique result in our study of edge states and topological properties of 1D p-wave SC. Finally, we show that the subgap states and their behaviour with $\phi$ depends heavily on $\mu$. We show this by calculating $E$ vs $\phi$ for the subgap states across junctions of BDI SC and compare the case when we have pure damped modes ($\mu<1/2$) to that when we have damped oscillating modes ($\mu>1/2$). Our work demonstrates that not only can we access different topological sectors in the coupled superconducting platform by using certain additional components, the behaviour of subgap states (and edge states) also changes considerably with the additional components. One interesting extension of our work would be to see the current behaviour across the phase boundaries between different regions of the phase diagram of the BDI class. Further, another natural extension of our work would be to study the additional effects of multi-terminal junctions, like in Ref.~\cite{xie, lucila, zazunov, julia} and interactions~\cite{dominguez, manisha, jelena2} to see the response of the two kinds of MBS we report in the BDI class in the coupled superconducting platform.

This work is supported by the NSF-FRHTP postdoctoral fellowship grant. S. R. acknowledges Ion Garate, Thomas Baker, Manisha Thakurathi, Abhiram Soori and DibyaKanti Mukerjee for useful discussions.


\section{Appendix} \label{appendix}

\subsection{Redefined symmetry operations} \label{redef_symm}
For cases, when more than one anti-unitary operator is present for a particular symmetry operation, the Hamiltonian needs to be block diagonalized into irreducible diagonal blocks and an effective operator needs to be identified for each such block to get the correct symmetry operation~\cite{tenfold}. For example, let us consider a 1D spinless p-wave BdG Hamiltonian in the presence of a s-wave pairing $\triangle_1 \sigma_y\tau_y$,
\begin{equation} \label{p+s_1}
\mathcal{H}(p)=\left(\frac{p^2}{2m}-\mu\right)\sigma_0\tau_z+\triangle_0 p\sigma_0\tau_x+\triangle_1 \sigma_y\tau_y,
\end{equation}
where,  $\sigma$ and $\tau$ are Pauli matrices for the spin and the particle-hole sector, $\triangle_0$ and $\mu$ are effective parameters, depending on the underlying heterostructure. The Nambu spinor is $\begin{pmatrix}
a^{\dagger}_{k,\uparrow} & a_{-k,\uparrow} & a^{\dagger}_{k,\downarrow} & a_{-k,\downarrow}
\end{pmatrix}$, and $\triangle_0$ is the magnitude of the p-wave pairing $\triangle_{\uparrow\uparrow} (=\triangle_{\downarrow\downarrow})$. We define the TR operator as an anti-unitary operator which does not mix the particle-hole sector. Whereas, the p-h operator is defined as an anti-unitary operator which rotates the particles into holes and holes into particle sector. It is not necessary for the TR operator to flip spin in all cases, which can be seen from the fact that the unperturbed 1D p-wave Kitaev model is TR invariant. TR operators, satisfying the condition  $\mathcal{T}\mathcal{H}(p)\mathcal{T}^{-1}=\mathcal{H}(-p)$, are $\mathcal{T}_1=\sigma_x\tau_z\mathcal{K}$ and $\mathcal{T}_2=\sigma_z\tau_z\mathcal{K}$. The Hamiltonian needs to be block diagonalized in the basis of $U_{\mathcal{T}_1}U^*_{\mathcal{T}_2}$, into $2\times2$ irreducible blocks, where $U_{\mathcal{T}}$ is the unitary operator for the corresponding TR operator $\mathcal{T}$. The above procedure allows us to diagonalize Eq.~\ref{p+s_1} into $\sigma_y=\pm1$ blocks:
\begin{equation} \label{p+s_block}
H_{1,2}=\left(\frac{p^2}{2m}-\mu\right)\sigma_0\tau_z+\triangle_0 p\sigma_0\tau_x\pm\triangle_1\tau_y
\end{equation}
The effective TR operator in each block is $\tau_z\mathcal{K}$. Similarly, there are two possible p-h operator for Eq.~\ref{p+s_1}, $\tau_x\mathcal{K}$ and $\sigma_y\tau_x\mathcal{K}$. However, in this case, due to the last term  $\triangle_1\tau_y$ in the block diagonalized form, neither of the two $\mathcal{C}$ operators is a symmetry. Thus, p-h symmetry is absent ($\mathcal{C}=0$) when a s-wave is present along with a p-wave of the given form. This class of Hamiltonian falls in the topologically trivial AI symmetry class (in $D=1$), with $\mathcal{T}^2=1$, $\mathcal{C}=0$ and $\mathcal{S}=0$,~\cite{tenfold}.

\subsection{Symmetry classes: Topologically trivial}
\subsection*{Case I: $\triangle_{\uparrow\uparrow}=\triangle_{\downarrow\downarrow}$}
Here we enlist the trivial classes and the perturbations that can lead to such classes for the first type of p-wave.
\subsubsection{AI class} \label{1_a1}
The AI symmetry class is a topologically trivial class in 1D, with $\mathcal{T}^2=1$, $\mathcal{C}=0$ and $\mathcal{S}=0$. The presence of the Zeeman term, s-wave term or combinations of both, can cause a transition from the BDI class of the unperturbed Hamiltonian in Eq.~\ref{ham1} to the AI class. We elaborate on the effect of each type of perturbation below.

\subsubsection*{\underline{$\mathbf{B_2.\hat{\sigma}\tau_0}$}}
Since the Hamiltonian in Eq.~\ref{ham1} does not have a spin operator, any magnetic field of the form ${\bf B}_2.\hat{\sigma}\tau_0$ can always be aligned along the z-direction, without any loss of generality. Thus, we can still break our Hamiltonian into two $2\times2$ irreducible blocks. This adds a constant term $B_2\sigma_z\tau_0$ in each $\sigma_z$ block, thus breaking the chiral symmetry. Thus, the MBS are no longer at $E=0$ and a transition occurs from the BDI class to the AI class.

\subsubsection*{\underline{$\mathbf{\triangle_1 \sigma_y\tau_y}$}}
With the first type of p-wave pairing, the presence of a s-wave term, like $\triangle_1 \sigma_y\tau_y$, is necessary for transition into the AI class, as already discussed in beginning of the paper.

\subsubsection*{\underline{$\triangle_1 \sigma_y\tau_y + B_2\sigma_y\tau_0$,  and,  $\triangle_1 \sigma_y\tau_y + B_1\sigma_y\tau_z$}}
Along with the spinless p-wave term and the s-wave term in Eq.~\ref{p+s_1}, we study the effect of two types of zeeman field $\mathbf{B_1\sigma.\tau_z}$ and $\mathbf{B_2\sigma.\tau_0}$:
\begin{eqnarray}\label{ham1_3}
\mathcal{H}^1_B(p) &=& \mathcal{H}_0(p)+\triangle_1 \sigma_y\tau_y+\mathbf{B}_1.\mathbf{\sigma} \mathbf{\tau_z} \nonumber\\
\mathcal{H}^2_B(p) &=& \mathcal{H}_0(p)+\triangle_1 \sigma_y\tau_y+\mathbf{B}_2.\mathbf{\sigma} \mathbf{\tau_0}
\end{eqnarray}
In the presence of $\triangle_1$, magnetic field along any $\hat{\sigma}$ cannot be aligned along $\sigma_z$ always. Thus, now the entire $4\times4$ Hamiltonian needs to be considered for symmetry analysis. However, when the zeeman field is along $\sigma_y$, the Hamiltonian can still be block diagonalized into $\sigma_y=\pm1$ blocks, and the symmetry analysis can be done for each block. This places $\mathcal{H}^1_B(p)$ and $\mathcal{H}^2_B(p)$  in the same class as $\mathcal{H}(p)$ in Eq.~\ref{p+s_1}. In each case, the effective TR operator is again $\tau_z\mathcal{K}$, p-h operator $\mathcal{C}=0$ and chiral operator $\mathcal{S}=0$.

\subsubsection*{\underline{$B_1\sigma_z\tau_z + B_2\sigma_z\tau_0$}}

For Hamiltonians with only spinless p-wave SC, $\triangle_{\uparrow\uparrow}=\triangle_{\downarrow\downarrow}$, in the simultaneous presence of both type of Zeeman fields:
\begin{equation} \label{ham_b1b2}
H({\bf p})=\left(\frac{p^2}{2m}-\mu\right)\tau_z +\triangle_0 p\tau_x + \mathbf{B}_1.\hat{\sigma}\tau_z +\mathbf{B}_2.\hat{\sigma}\tau_0
\end{equation}
(where, $\mathbf{B}_1$ and $\mathbf{B}_1$ are the two different Zeeman field), the AI class is again possible. There are two possible TR operators $\tau_z\mathcal{K}$ and $\sigma_z\tau_z\mathcal{K}$, with effective operator in each irreducible block being $\tau_z\mathcal{K}$. However, the p-h $\mathcal{C}$ and chiral operator $\mathcal{S}$ are still absent.

\begin{table*}[h]
	\centering
	\caption{Summary table for perturbations giving AI class for Case I ($\triangle_0 p\sigma_0 \tau_x$)}
	\setlength{\tabcolsep}{13pt}
	\begin{tabular}{l | c | c | c }
		Perturbation & $\mathcal{T}$ & $\mathcal{C}$ & $\mathcal{S}$ \\
		\hline \hline
		$\mathbf{B}_2\hat{\sigma}\tau_0$ & 
		\begin{minipage}{2.5cm}
			$\tau_z\mathcal{K}$\\
			$\mathcal{T}^2=1$
		\end{minipage}
		& 
		\begin{minipage}{2.5cm}
			$\sigma_y\tau_x\mathcal{K},\
			\tau_x \mathcal{K}
			$ \\
			$\mathcal{C}_{eff}=0$
		\end{minipage}
		& 0 \\ 
		\\
		\hline
		$\triangle_1 \sigma_y\tau_y$ & 
		\begin{minipage}{2cm}$
			\tau_z\mathcal{K}
			$\\
			$\mathcal{T}^2=1$
		\end{minipage}
		& 
		\begin{minipage}{2cm}
			$ \tau_x\mathcal{K},\
			\sigma_y\tau_x\mathcal{K}
			$ \\
			$\mathcal{C}_{eff}=0$
		\end{minipage}
		& 
		0
		\\ 
		\\
		\hline
		$\triangle_1 \sigma_y\tau_y+B_2\sigma_y\tau_0$,\\
		$\triangle_1 \sigma_y\tau_y+B_1\sigma_y\tau_z$ & 
		\begin{minipage}{1.5cm}$
			\tau_z \mathcal{K},
			$
			$\mathcal{T}^2=1$
		\end{minipage}
		& 
		0
		& 
		0
		\\ 
		\\
		\hline
		$B_1\sigma_z\tau_z + B_2\sigma_z\tau_0$& 
		\begin{minipage}{2cm}
			$\tau_z \mathcal{K},\
			\sigma_z\tau_z \mathcal{K}
			$
			$\mathcal{T}_{eff}=\tau_z\mathcal{K}$\\
			$\mathcal{T}^2=1$
		\end{minipage}
		& 
		0
		& 
		0
		\\ 
		\\
		\hline
	\end{tabular}
	
	\label{table1}
\end{table*}

\subsubsection{CI class} \label{1_c1}
The CI class is another topologically trivial class in 1D with $\mathcal{T}^2=1$, $\mathcal{C}^2=-1$ and $\mathcal{S}=1$. The p-wave Hamiltonian in Eq.~\ref{ham1} can be transferred to the CI class in the presence of perturbations like the s-wave pairing along with the Zeeman term $\mathbf{B_2}$.

\subsubsection*{\underline{$\triangle_1 \sigma_y\tau_y + B_2 \sigma_x\tau_0$}}
With the above perturbation over the p-wave case in Eq.~\ref{ham1}, the TR operator is $\sigma_x\tau_z\mathcal{K}$, p-h operator is $\sigma_y\tau_x\mathcal{K}$ and the chiral operator is $\sigma_z\tau_y$, rendering it to be in the CI class.

\subsubsection*{\underline{$\triangle_1 \sigma_y\tau_y + B_2 \sigma_z\tau_0$}}
Here again, the above perturbation places the Hamiltonian in Eq.~\ref{ham1} in the topologically trivial CI class, with $\mathcal{T}^2=1$, $\mathcal{C}^2=-1$ and $\mathcal{S}=1$. The TR operator is $\sigma_z\tau_z\mathcal{K}$, p-h operator is $\sigma_y\tau_x\mathcal{K}$ and the chiral operator is $\sigma_z\tau_y$. 

\begin{table*}[h]
	\centering
	\caption{Summary table for perturbations giving CI class for Case I ($\triangle_0 p\sigma_0 \tau_x$)}
	\setlength{\tabcolsep}{13pt}
	\begin{tabular}{l | c | c | c }
		Perturbation & $\mathcal{T}$ & $\mathcal{C}$ & $\mathcal{S}$ \\
		\hline \hline
		$\triangle_1 \sigma_y\tau_y+B_2\sigma_x\tau_0$ & 
		\begin{minipage}{2.5cm}
			$\sigma_x\tau_z\mathcal{K}$\\
			$\mathcal{T}^2=1$
		\end{minipage}
		& 
		\begin{minipage}{2.5cm}
			$\sigma_y\tau_x\mathcal{K}
			$ \\
			$\mathcal{C}^2=-1$
		\end{minipage}
		& \begin{minipage}{2.5cm}
			$\sigma_z\tau_y\mathcal{K}
			$ \\
			$\mathcal{S}=1$
		\end{minipage} 
		\\ 
		\\
		\hline
		$\triangle_1 \sigma_y\tau_y+B_2\sigma_z\tau_0$ & 
		\begin{minipage}{2cm}$
			\sigma_z\tau_z\mathcal{K}
			$\\
			$\mathcal{T}^2=1$
		\end{minipage}
		& 
		\begin{minipage}{2cm}
			$
			\sigma_y\tau_x\mathcal{K}
			$ \\
			$\mathcal{C}^2=-1$
		\end{minipage}
		& 
		\begin{minipage}{2.5cm}
			$\sigma_z\tau_y\mathcal{K}
			$ \\
			$\mathcal{S}=1$
		\end{minipage} 
		\\ 
		\\
		\hline
	\end{tabular}
	
	\label{table2}
\end{table*}

\subsection{Case II: $\triangle_{\uparrow\uparrow}=-\triangle_{\downarrow\downarrow}$}
Below we list the trivial classes and the perturbations that can lead to such classes for the second type of p-wave.

\subsubsection{A class} \label{2_a}
The A symmetry class is a topologically trivial class in 1D with $\mathcal{T}$, $\mathcal{C}$ and $\mathcal{S}$ all equal to zero. Combination of the s-wave pairing with the $\mathbf{B}_1$ Zeeman field can cause a transition from the BDI class to the A class in the unperturbed p-wave Hamiltonian in Eq.~\ref{ham2}.

\subsubsection*{\underline{$\triangle_1 \sigma_y\tau_y+B_1 \sigma_x \tau_z$}}
A perturbation of the above form on Eq.~\ref{ham2}, will generate the A symmetry class. Even though for the $4\times4$ Hamiltonian there are two possible p-h operators $\tau_x \mathcal{K}$ and $\sigma_x\tau_y\mathcal{K}$, after block diagonalization, each irreducible block becomes:
\begin{equation}
H_{1,2}= -\left(\frac{p^2}{2m}-\mu\right)\tau_z-\triangle_0 p \tau_x+\triangle_1\tau_x-B_1 \tau_0,
\end{equation}
and none of the above p-h operators is a symmetry. This is another example which shows that even though the entire $4\times4$ Hamiltonian has more than one possible symmetry operator, when block diagonalized into an irreducible form the symmetry might be absent in each block.

\subsubsection{C class} \label{2_c}
The C symmetry class is again a topologically trivial class in $D=1$ with $\mathcal{T}=0$, $\mathcal{C}^2=-1$ and $\mathcal{S}=0$. Here again, perturbations like the s-wave term and the Zeeman terms $\mathbf{B}_1$ and $\mathbf{B}_2$ on the unperturbed p-wave case, can cause a transition to the C class.

\subsubsection*{\underline{$\triangle_1 \sigma_y\tau_y+B_2 \sigma_z \tau_0$ and $\triangle_1 \sigma_y\tau_y+B_1 \sigma_y \tau_z$}}
For each of the above perturbations on Eq.~\ref{ham2}, the symmetry operations follow the above conditions, rendering the system topologically trivial. In both cases, the entire $4\times4$ Hamiltonian is irreducible, with the p-h operator being $\sigma_x\tau_y\mathcal{K}$ in each case.

\subsubsection{AI class} \label{2_a1}
Another topologically trivial class which can be accessed by the 1D spinless p-wave SC is the AI class, as seen in the previous section Sec.~\ref{1_a1}. The symmetry conditions are, $\mathcal{T}^2=1$, $\mathcal{C}=0$ and $\mathcal{S}=0$. Note that unlike in Sec.~\ref{1_a1}, s-wave perturbations will not cause a transition to the AI class. For this second type of p-wave, only combinations of the Zeeman fields induce a transition to this topologically trivial class.

\subsubsection*{\underline{$B_2 \sigma_z \tau_0$}}
With the unperturbed Hamiltonian in Eq.~\ref{ham2}, $B_2 \sigma_z \tau_0$ adds as a constant term in each diagonal block, thus breaking chiral symmetry in each. The TR operator is $\tau_z\mathcal{K}$ and the effective p-h operator in each block is 0. Thus, $\mathcal{T}^2=1$, $\mathcal{C}=0$ and $\mathcal{S}=0$ for this perturbation, placing it in the topologically trivial AI class.

\subsection*{\underline{$B_1 \sigma_x \tau_z$}}
With the above perturbation, the possible TR operators are $\sigma_x\mathcal{K}$ and $\tau_z\mathcal{K}$, with the effective $\mathcal{T}$ for each block being $\tau_z\mathcal{K}$. Similarly, possible p-h operators are $\tau_x\mathcal{K}$ and $\sigma_x\tau_y\mathcal{K}$, but the effective $\mathcal{C}$ is 0. Chiral symmetry $\mathcal{S}$ is also zero for the above perturbation.

\subsubsection*{\underline{$B_1 \sigma_y \tau_z$}}
Even though $\sigma_x\mathcal{K}$ is the TR operator for the entire $4\times4$ Hamiltonian with $B_1 \sigma_y \tau_z$, the effective $\mathcal{T}$ in each irreducible block is $\tau_z\mathcal{K}$. Similarly, possible p-h operators are $\sigma_x\tau_y\mathcal{K}$ and $\sigma_z\tau_x\mathcal{K}$, but the effective $\mathcal{C}$ is 0. Chiral symmetry, again, is zero for the above perturbation.

\subsubsection*{\underline{Combination of $\mathbf{B_1 \sigma \tau_z}$ and $\mathbf{B_2 \sigma \tau_0}$}}
Certain combinations of the two Zeeman fields in Eq.~\ref{ham2} (present simultaneously on the p-wave SC) also give the AI class. 
\begin{itemize}
	\item $B_1 \sigma_z \tau_z+B_2 \sigma_z \tau_0$: The possible TR operators are $\tau_z\mathcal{K}$ and $\sigma_z\tau_z\mathcal{K}$, with $\mathcal{T}_{eff}=\tau_z\mathcal{K}$ in each block. $\mathcal{C}$ and $\mathcal{S}$ are 0.
	
	\item $B_1 (\sigma_x+\sigma_z)\tau_z+B_2 \sigma_z \tau_0$ and $B_1 (\sigma_x+\sigma_z)\tau_z+B_2 \sigma_x \tau_0$: For each combination, $\mathcal{T}$ is $\tau_z\mathcal{K}$ for the entire $4\times4$ Hamiltonian. $\mathcal{C}$ and $\mathcal{S}$ are again 0.
	
\end{itemize}

\begin{table*}[h]
	\centering
	\caption{Summary table for perturbations giving AI class for Case II ($\triangle_0 p\sigma_z \tau_x$)}
	\setlength{\tabcolsep}{13pt}
	\begin{tabular}{l | c | c | c }
		Perturbation & $\mathcal{T}$ & $\mathcal{C}$ & $\mathcal{S}$ \\
		\hline \hline
		$B_2\sigma_z\tau_0$ & 
		\begin{minipage}{2.5cm}
			$\tau_z\mathcal{K}$\\
			$\mathcal{T}^2=1$
		\end{minipage}
		& 
		0
		&
		0
		\\ 
		\\
		\hline
		$B_1\sigma_x\tau_z$  & 
		\begin{minipage}{2cm}$
			\sigma_x\mathcal{K},\
			\tau_z\mathcal{K}
			$\\
			$\mathcal{T}_{eff}=\tau_z\mathcal{K}$\\
			$\mathcal{T}^2=1$
		\end{minipage}
		& 
		\begin{minipage}{2cm}
			$ \tau_x\mathcal{K},\
			\sigma_x\tau_y\mathcal{K}
			$ \\
			$\mathcal{C}_{eff}=0$
		\end{minipage}
		& 
		0
		\\ 
		\\
		\hline
		$B_1\sigma_y\tau_z$ & 
		\begin{minipage}{1.5cm}$
			\sigma_x\mathcal{K},
			$\\
			$\mathcal{T}_{eff}=\tau_z \mathcal{K}$\\
			$\mathcal{T}^2=1$
		\end{minipage}
		& 
		\begin{minipage}{2cm}
			$ \sigma_z\tau_x\mathcal{K},\
			\sigma_x\tau_y\mathcal{K}
			$ \\
			$\mathcal{C}_{eff}=0$
		\end{minipage}
		& 
		0
		\\ 
		\\
		\hline
		$B_1\sigma_z\tau_z + B_2\sigma_z\tau_0$& 
		\begin{minipage}{2cm}
			$\tau_z \mathcal{K},\
			\sigma_z\tau_z \mathcal{K}
			$\\
			$\mathcal{T}_{eff}=\tau_z \mathcal{K}$\\
			$\mathcal{T}^2=1$
		\end{minipage}
		& 
		0
		& 
		0
		\\ 
		\\
		\hline
		$B_1(\sigma_x+\sigma_z)\tau_z + B_2\sigma_z\tau_0$\\ 
		$B_1(\sigma_x+\sigma_z)\tau_z + B_2\sigma_x\tau_0$&
		\begin{minipage}{2cm}
			$\tau_z \mathcal{K}
			$\\
			$\mathcal{T}^2=1$
		\end{minipage}
		& 
		0
		& 
		0
		\\
		\\
		\hline
	\end{tabular}
	\label{table4}
\end{table*}

\subsection{Analytical derivation of subgap states in the zero barrier limit} \label{analytical_subgap}

\begin{figure*}[h]
\centering
\text{Analytical solutions for subgap states}\par\medskip
\centerline{\includegraphics[width=0.46\linewidth]{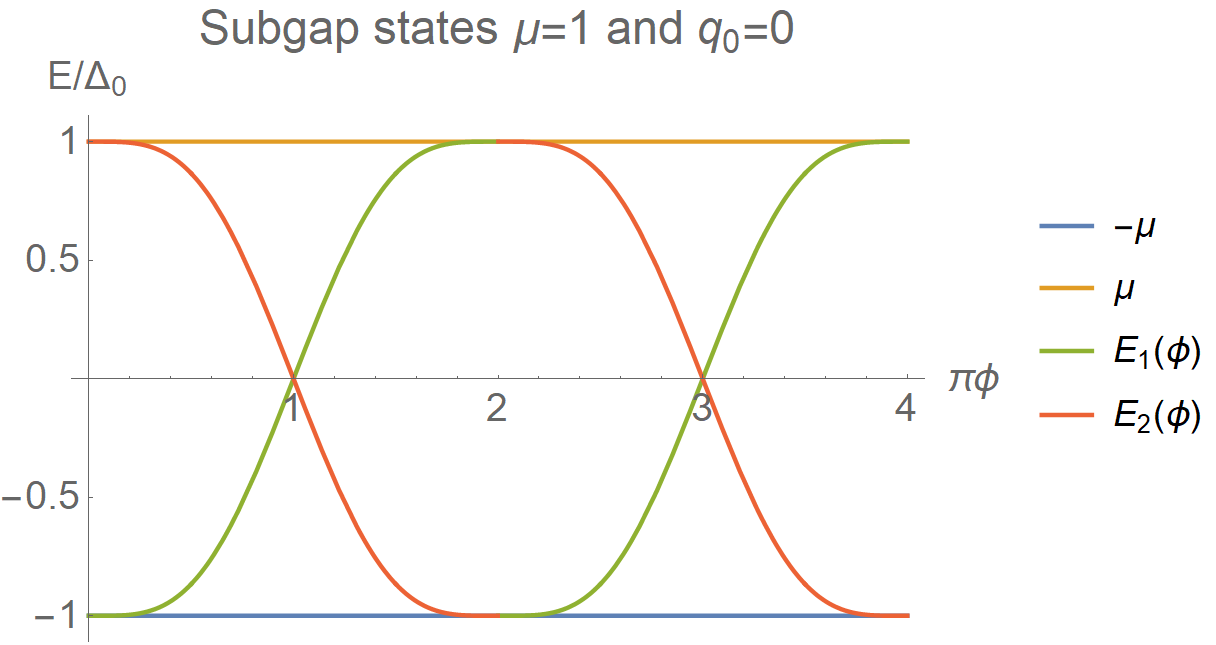}~
\includegraphics[width=0.5\linewidth]{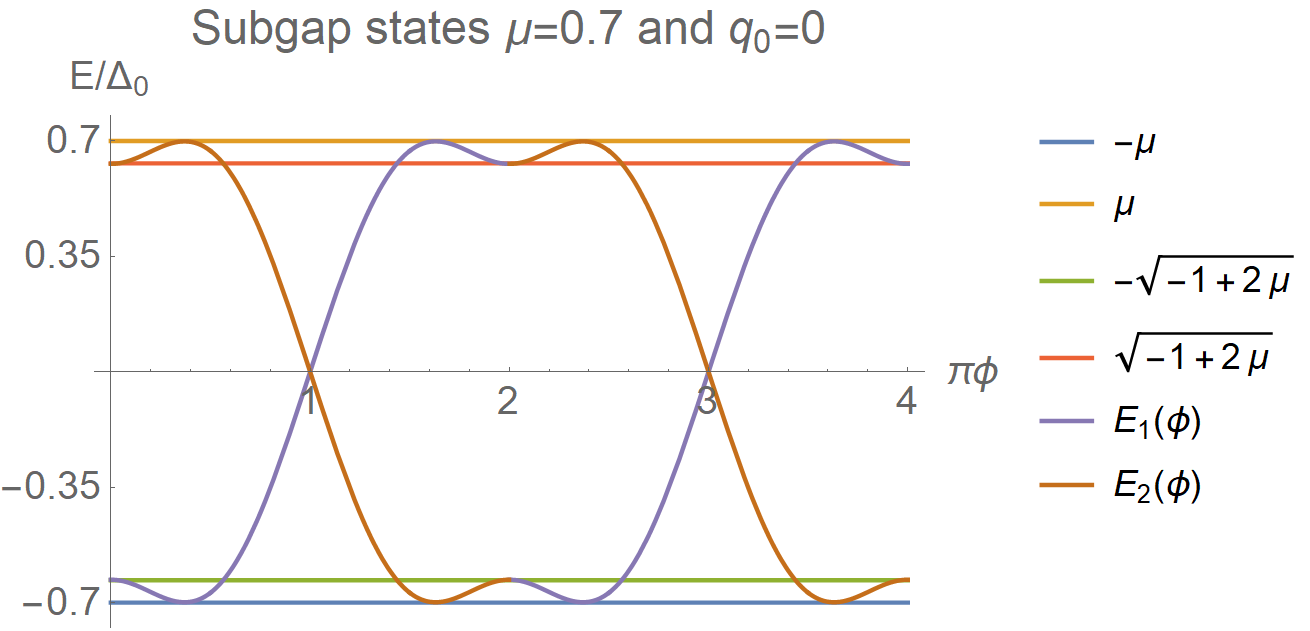}}
 \caption{Analytical solutions for subgap states in the zero barrier limit, across junctions of BDI class, with $\triangle_0=1$. (Left) For $\mu=1$. (Right) $\mu=0.7$.}
\label{analytical_zerobarrier}
\end{figure*}

In the zero barrier limit, one can analytically solve for the solutions of the subgap states from the secular equation in Eq.~\ref{boundarycond}, where the determinant $f(E, \mu, \triangle_0, q_0, \phi)=0$. In this limit, the solutions for subgap states are at $E=\pm\mu$, $E=\pm \sqrt{-1+2\mu}$, $E_1(\phi) = -\frac{1}{2} i e^{-\frac{i}{2} \pi  \phi }  \cos \left(\frac{\pi  \phi }{2}\right) \sqrt{-8 \mu  e^{i \phi }+2 e^{i \phi }+e^{2 i \phi }+1}$ and $E_2(\phi) = \frac{1}{2} i e^{-\frac{i}{2} \pi  \phi }  \cos \left(\frac{\pi  \phi }{2}\right) \sqrt{-8 \mu  e^{i \phi }+2 e^{i \phi }+e^{2 i \phi }+1}$. We plot the solutions for $\mu=1$ and $\mu=0.7$ in Fig.~\ref{analytical_zerobarrier}. When $\mu=1$, the solutions $E=\pm \mu$ and $E=\pm \sqrt{-1+2\mu}$ become equal. When $\mu \neq 1$, the solutions become different, and additional states start to appear for $|\mu|>E>|\sqrt{-1+2\mu}|$. However, on analyzing the nature of the wavefunctions of the subgap states, one can get the physically valid solutions. Also, on careful investigation of the zeroes of the determinant , we can further say that $E_{c}=\pm \sqrt{-1+2\mu}$ are not solutions for all $\phi$'s but is only true when $\pi \phi =2 n$, for $n$ integer. Hence, when we numerically solve for the solutions in Fig.~\ref{mu_bdI}, we do not get the solutions at $E_{c}=\pm \sqrt{-1+2\mu}$ at all $\phi$'s. We also get discontinuities at $E_{c}$ because of the changing nature of the subgap states and the determinant.

\end{document}